\documentclass[aps,pra,twocolumn,superscriptaddress,amsmath,amssymb,showpacs,notitlepage]{revtex4-1}
\pdfoutput=1

\usepackage{color,xcolor}
\usepackage{graphicx}
\usepackage{bm}
\usepackage{hyperref}
\usepackage{mathtools}
\usepackage{subfigure}
\usepackage{wrapfig}
\usepackage{braket}
\usepackage{array}
\usepackage{bibentry}
\usepackage{boxhandler}

\makeatletter
\newcommand\mathcircled[1]{%
    \mathpalette\@mathcircled{#1}%
}
\newcommand\@mathcircled[2]{%
    \tikz[baseline=(math.base),outer sep=auto]{\node[draw,circle,inner
    sep=1pt]
    (math) {$#1#2$};
    \path (math.north)--++(0,1pt);
    \path (math.south)--++(0,-1pt);}%
}
\makeatother
\graphicspath{ {./images/} }
\begin{document}


\title{Entanglement generation in a quantum network at distance-independent rate}%

\author{Ashlesha Patil}
\email{ashlesha@email.arizona.edu}
\affiliation{College of Optical Sciences, University of Arizona, 1630 East University Boulevard, Tucson, AZ 85721.}
 \author{Mihir Pant}
 \affiliation{Massachusetts Institute of Technology, 32 Vassar Street, Cambridge MA 02139}
  \author{Dirk Englund}
 \affiliation{Massachusetts Institute of Technology, 32 Vassar Street, Cambridge MA 02139}
 \author{Don Towsley}
 \affiliation{ College of Information and Computer Sciences, University of Massachusetts, Amherst, MA 01002.}
\author{Saikat Guha}
\affiliation{College of Optical Sciences, University of Arizona, 1630 East University Boulevard, Tucson, AZ 85721.}

\begin{abstract}
We develop a protocol for entanglement generation in the quantum internet that allows a repeater node to use $n$-qubit Greenberger-Horne-Zeilinger (GHZ) projective measurements that can fuse $n$ successfully-entangled {\em links}, i.e., two-qubit entangled Bell pairs shared across $n$ network edges, incident at that node. Implementing $n$-fusion, for $n \ge 3$, is in principle not much harder than $2$-fusions (Bell-basis measurements) in solid-state qubit memories. If we allow even $3$-fusions at the nodes, we find---by developing a connection to a modified version of the site-bond percolation problem---that despite lossy (hence probabilistic) link-level entanglement generation, and probabilistic success of the fusion measurements at nodes, one can generate entanglement between end parties Alice and Bob at a rate that stays constant as the distance between them increases. We prove that this powerful network property is not possible to attain with any quantum networking protocol built with Bell measurements and multiplexing alone. We also design a two-party quantum key distribution protocol that converts the entangled states shared between two nodes into a shared secret, at a key generation rate that is independent of the distance between the two parties.

\end{abstract}

\maketitle


\section{\label{sec:intro}Introduction}
The Quantum Internet will provide the service of generating shared entanglement of different kinds, between distant end-user pairs and groups, on demand, and at high speeds. The entanglement generation rate between two nodes decays linearly with the transmissivity $\eta$ of the channel connecting them, which turns into an exponentially rate-vs.-distance decay over optical fiber, since $\eta = e^{-\alpha L}$ for a length-$L$ fiber~\cite{Takeoka2014-bh}. The maximum attainable rate is $-\log(1-\eta) \approx 1.44 \eta$, for $\eta \ll 1$, ebits (pure Bell states shared between two parties) per transmitted optical mode~\cite{Pirandola2017-ej}. {\em Quantum repeaters} need to be inserted along the length of the optical channel in order to circumvent this rate-vs.-loss limit~\cite{Guha2015-yj, Muralidharan2016-mn, Pant2017-mp}. In~\cite{pirandola2019end}, the ultimate capacity of entanglement generation between two nodes was found, assuming the network nodes were fully-equipped quantum processors. A consequence of this result was that multipath routing can improve entanglement rates over that attainable by routing along one path.

There is a wide variety of repeater and router protocols being researched using practical devices and systems, most of which use Bell state measurements (BSMs) as a building block. BSM is a two-qubit destructive measurement that can fuse two entangled links (each entangled link being a two-qubit Bell state, each shared across a network edge) incident at a node, into one entangled link over a two-hop path. For a linear chain of repeater nodes, where each repeater is equipped with quantum memories and employs BSMs and switches, the entanglement rate outperforms what can be attained with a direct connection connecting the communicating end parties, but the rate still decays exponentially with distance, i.e., $R \sim e^{-s \alpha L}$, with $s<1$~\cite{Guha2015-yj}.


Various genres of quantum repeaters and associated error-correction codes are under investigation~\cite{Muralidharan2016-mn}. For the purposes of our paper, we will consider the following simple model, and show a surprising result---that the end-to-end entanglement rate between two uses Alice and Bob remains constant with increasing distance when network nodes are able to measure more than two qubits in a joint projective measurement. In each time slot, each network edge attempts to establish an entangled link: a Bell state of two qubits, each residing in a quantum memory at nodes on either end of the link. In every time slot, each link is established successfully, i.i.d., with probability $p$ proportional to the transmissivity of the optical link. Subsequently, each node, based on local link-state information (i.e., which neighboring links succeeded in that time slot), and knowledge of the location of the communicating parties Alice and Bob, decides which pairs of successful links to fuse. The two qubits that are fused with a BSM at a node are destroyed in the measurement process, while creating an entangled (Bell) state among the two qubits at the far ends of the two links, thus creating a $2$-hop entangled link traversing two network edges. A fusion attempt succeeds with probability $q$. It was shown recently that with a simple distance-vector fusion rule, the achievable entanglement generation rate exceeds what is possible with a fusion schedule along a pre-determined single shortest path connecting Alice and Bob~\cite{pant2019routing}. Despite this rate advantage from multipath entanglement routing, the rate decays exponentially with the distance $L$ between Alice and Bob, for any value of $p$ or $q$ less than $1$. Interestingly, this exponential scaling of rate with multipath routing, is at odds with the network's capacity proven in~\cite{pirandola2019end}, according to which the end-to-end rate does not even depend upon the distance between communicating parties. The protocol we develop in this paper addresses this gap.

In this paper, we develop a protocol that allows nodes to use $n$-qubit Greenberger-Horne-Zeilinger (GHZ) projective measurements, i.e., $n$-fusions, that can fuse $n$ successful links at a node. When $n=2$, the nodes implement a two-qubit BSM. For $n=1$, the nodes implement a single-qubit Pauli measurement. Implementing $n$-fusion, for $n \ge 3$ is in principle no harder than $2$-fusions (Bell measurements) in qubit memories, e.g., color centers in diamond~\cite{Bhaskar2020-wi}, and trapped-ion quantum processors~\cite{Brown2016-dk}. We take the success probability of an $n$-fusion attempts as $q$. We report a surprising result: if we allow even $3$-fusions at the repeater nodes, there is a non-trivial regime of $(p, q)$ where our protocol generates entanglement at a rate that stays constant with $L$. We prove this is not possible with any quantum network protocol that only uses Bell measurements (see Section~\ref{app:2GHZ}). Our protocol only uses local link state knowledge, but requires a single-round of classical communications that adds to the latency of the protocol (but, does not affect the rate). 

Finally, we develop a quantum key distribution (QKD) protocol that allows a pair of users Alice and Bob, situated in a network, to sift (two-party) secret keys starting from a pre-shared $m+n$-qubit Greenberger-Horne-Zeilinger (GHZ) state, $m$ qubits of which are held by Alice and $n$ by Bob. It is an extension of the BBM’92 protocol~\cite{Bennett1992-bm}, a simplification of the E'91 protocol~\cite{Ekert1991-jo}, which relies on shared Bell states and measurements by Alice and Bob in a matching pair of bases. Using our above described quantum network protocol that employs $n$-fusions at nodes, we thus have devised a QKD protocol over a quantum network whose secret-key generation rate is constant with increasing distance between communicating parties, despite lossy channel segments between nodes and probabilistic successes of the $n$-fusions at nodes.

In Section \ref{sec:quantum operations}, we discuss the elementary multi-qubit projective measurements used in our protocol. Section \ref{sec:protocol} describes the entanglement distribution protocol. We also map the problem of distributing entanglement over a quantum network to a mixed percolation problem studied in classical statistical mechanics. We discuss the origin of distance-independence of the shared entanglement rate, along with numerical calculations of the rate and comparisons with capacity, in Section \ref{sec:results} and the improved variation of the protocol in Section \ref{sec:improved 3-GHZ}. Section \ref{sec:QKD} describes the QKD protocol using GHZ states. We conclude in Section~\ref{sec:conclusion} by summarizing the results and discussing open questions that can be studied as immediate extensions and applications of the proposed protocol.

\section{Fusing entanglement using GHZ-state projections}\label{sec:quantum operations}
We use entanglement-swapping operations, namely, Bell State Measurements (BSMs) and $n$-qubit GHZ projections at network nodes, for routing entanglement in a quantum network. An $n$-qubit GHZ projection is a von Neumann projective measurement, that projects the $n$ measured qubits into one of the ($2^n$) mutually-orthogonal $n$-qubit GHZ states, thereby producing a (random) $n$-bit classical measurement result. The well-known BSM is a $2$-qubit GHZ projection. Entanglement swapping at a quantum (repeater) node extends the range of entanglement by {\em fusing} two Bell states shared across two adjacent edges of the network.

\begin{figure}[h]
    \centering
    \includegraphics{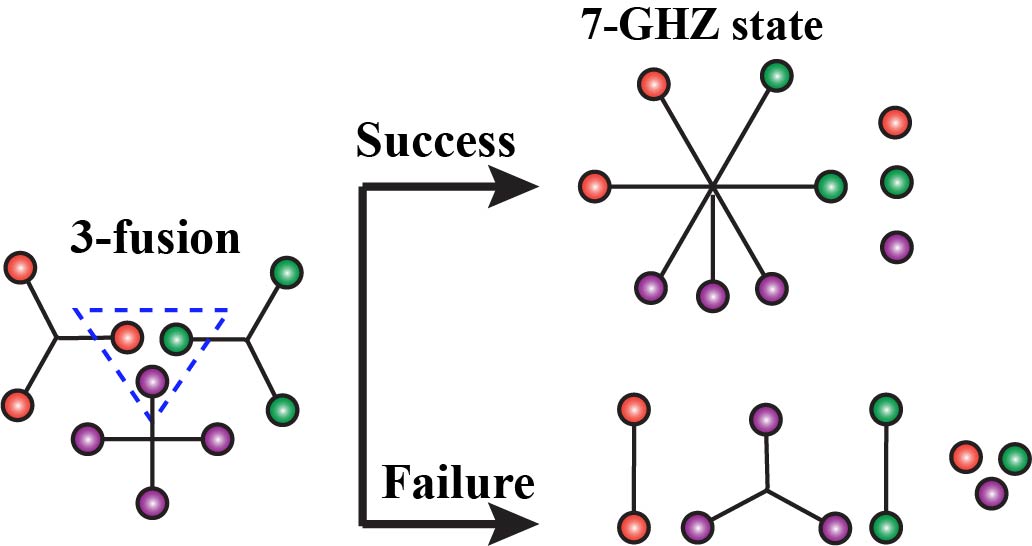}
    \caption{$3$-fusion on two 3-GHZ states and a 4-GHZ state. Fusion success creates a $7$-GHZ state and failure performs X-basis measurements on the fused qubits, resulting in two $2$-GHZ states (Bell pairs) and a 3-GHZ state. Even though the states are represented using graphs, they are not cluster states.}
    \label{fig:GHZ_proj}
\end{figure}

We refer to $n$-qubit stabilizer states~\cite{Poulin2005-ic} with stabilizer generators of the form $\{(-1)^{g_1}X_1X_2\dots X_n, (-1)^{g_2}Z_1Z_2, (-1)^{g_3}Z_1Z_3, \dots, \\ (-1)^{g_n}Z_1Z_n\}$, $g_i \in \{0,1\}$ as $n$-GHZs, which includes the case of $n=2$ i.e., Bell states. $X_i$ and $Z_i$ are single-qubit Pauli operators for the $i$-th qubit. We use the (unconventional) notation of an $n$-star graph to represent an $n$-GHZ. This is not a star-topology commonly-known cluster state~\cite{Nielsen2006-zi}. Furthermore, we refer to a projective measurement onto the $n$-GHZ basis as a ($n$-qubit) {\em fusion}. The {\em size} of an $n$-qubit GHZ state is $n$. An $n$-fusion on a set of GHZ states of size ${m_1, m_2,....m_n}$ results in a single GHZ state of size $\sum_{i=1}^n m_i - n$, obtained by removing the qubits that are fused from the original set of qubits and coalescing all the unmeasured qubits into a single GHZ state, as shown in Fig.~\ref{fig:GHZ_proj}.

\begin{figure}[ht]
    \centering
    \includegraphics{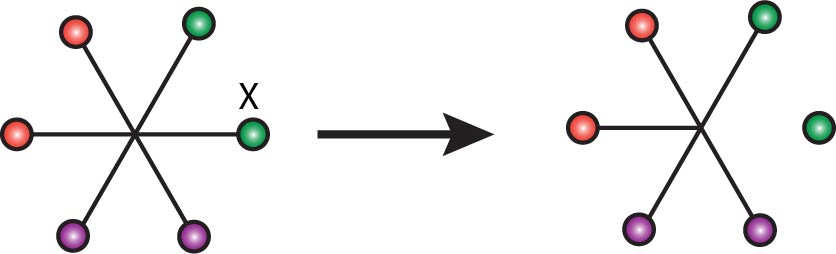}
    \caption{Measuring a qubit in X-basis removes it from the $n$-GHZ state.  Even though the states are represented using graphs, they are not cluster states.}
    \label{fig:Xmeasurement}
\end{figure}

Depending upon the choice of quantum memory and processor hardware at the quantum repeater node, fusion operations may be probabilistic. We model the result of a failed fusion attempt as performing an $X$-basis measurement on all qubits that were used as part of the fusion, as shown in Fig.~\ref{fig:GHZ_proj}. Measuring a qubit of an $n$-GHZ state in the Pauli-$X$ basis results in a $(n-1)$-GHZ state, unentangled with the measured qubit, as shown in Fig.~\ref{fig:Xmeasurement}. 

\section{The Protocol}\label{sec:protocol}
In this paper, we study two kinds of quantum networks: a two-dimensional square-grid, and a configuration-model random graph with a given node-degree distribution~\cite{Newman2001-fz}. First, let us consider a square-grid graph. Each node is a quantum repeater (blue circles in Fig.~3a) with four quantum memories (black dots in Fig.~3a) associated with each neighboring edge. Each repeater is either a ``consumer" of entanglement i.e., Alice and Bob, or a ``helper" i.e., they help to establish entanglement between the consumer nodes. In the first time step, each network edge attempts to establish an entangled link: a Bell state of two qubits, each residing in a quantum memory at nodes on either end of the link. Each link is established successfully, i.i.d., with probability $p$, which is proportional to the transmissivity of the respective optical link~\cite{Takeoka2014-bh,Pirandola2017-ej}. The repeater nodes have only local link-state knowledge, i.e., a repeater knows the success-failure outcomes at each time slot of its own link generation attempts (across its neighboring edges). Each repeater is also aware whether it is a consumer or a helper node, knows the overall network topology, and the location of the consumer nodes (if it is a helper node). In the next time step, all helper nodes that have more than one successfully-created link, attempt fusions on the qubits held in their respective quantum memories tied to a subset of those successful links. The fusion success probability is taken to be $q$. A successful fusion at a repeater creates a Bell pair or a GHZ state shared between a subset of its neighbours. If a helper node has only one link success in a time slot, it performs an X-basis measurement on the corresponding locally-held qubit, which unentangles and dissociates that qubit from any others in the network. 

We consider three protocols for the square-grid network which differ in the operations available at repeater nodes, and allow for different entanglement generation rates: (1) the $4$-GHZ protocol, (2) the $3$-GHZ protocol and (3) the $3$-GHZ brickwork protocol 

In the $4$-GHZ protocol, a repeater performs a fusion on all locally-held qubits successfully entanlged with neighboring nodes at each time step. Hence, the largest measurement in such a protocol is a 4-GHZ measurement, which is done when all 4 links are successfully created. In a time step when only $3$ or $2$ links are successful, a 3-GHZ measurement or a 2-GHZ (i.e., Bell) measurement is performed. If only one link is successful, the corresponding qubit is measured in the X-basis. In the 3-GHZ protocol, the maximum size of the GHZ projection allowed is limited to 3, which may be imposed due to hardware constraints. If the number of successful neighboring links of a helper node is less than or equal to three, the repeater performs a fusion between the corresponding qubits, i.e., behaves the same as the 4-GHZ protocol. However, if four neighboring links are successful, the repeater randomly chooses three qubits and performs a fusion on them. It performs an X measurement on the fourth remaining qubit if this happens. Every helper node sends its local link state knowledge, fusion success outcomes, and X-basis measurement outcomes to the consumers Alice and Bob using a classical communication overlay channel. This classical communication time determines the overall latency of the entanglement generation protocol, but the entanglement rate is determined by the rate at which each entangled link is attempted across each network edge.

It is important to note that {\em all} Bell state measurements, GHZ projections and Pauli X-basis measurements across the entire network are performed during the same time step. This is allowed because all of these operations and measurements commute with one another. At the end of this step, the consumers obtain (potentially more than one) shared $m$-qubit GHZ state(s) with a probability that depends on the network topology, $p$, $q$, and which of the two protocols described above is used. 

We discuss the rules for the Brickwork protocol in section~\ref{sec:brickwork}, which instead of being fully randomized as above, imposes some additional structure on which fusions to attempt, and can outperform the $3$-GHZ protocol in certain regimes.

We also study the $n$-GHZ protocol for a random graph network, with an arbitrary node degree distribution $p_k$. 
Here, $p_k$ is the probability that a randomly chosen node has degree $k$. In other words, it is the probability that a randomly chosen quantum repeater node has $k$ edges. In an $n$-GHZ protocol, each repeater performs $m$-GHZ projections for fusions where $m=\min(n, \text{no. of successful links at the repeaters})$ i.e., repeaters can perform up to $n$-fusions. For the $n$-GHZ protocol over a random network, if a degree $k$ helper node has $l$ successful links in a time slot such that $l\leq n$, it performs an $l$-GHZ fusion. If $l>n$, it performs an $n$-GHZ fusion on the $n$ qubits corresponding to $n$ randomly chosen links (of the $l$). The remaining steps are same as the 3- and 4-GHZ protocols described above for the square grid. 

Immediately after the time slot when all helper nodes perform their measurements and sends, via unicast communications, the requisite classical communication to the consumer nodes, the network edges re-attempt entanglement generation in the next time step, and the helper nodes again make their measurements based on the protocol described above using local link state information, until the end of the protocol's duration. 
The length of each time step determines the rate of the protocol, whereas the classical communication time determines the latency. Consumers hold on to all their qubits for the time required to receive the classical information regarding the results of measurements made during a specific time slot from every helper node in the network. They use the local link state knowledge from the helpers to determine which one of their qubits (from the corresponding time slot) are part of a shared entangled state held between Alice and Bob. In each time slot, Alice and Bob generate 0, 1, 2, 3 or 4 shared GHZ states. Each of those shared GHZ states could have more than $2$ qubits. For example, Alice and Bob could generate one $3$-qubit GHZ state two of whose qubits are held by Alice and one by Bob, and one $2$-qubit GHZ (i.e., Bell) state one qubit of which is with Alice and the other with Bob.

At this point, Alice and Bob can use their shared entangled state for a quantum information processing protocol, e.g., QKD, entanglement enhanced sensing, or distributed quantum computing implemented by a teleported gate. If the protocol requires a particular $n$-GHZ state as a resource, it is always possible for Alice and Bob to correct the state by applying local unitary operations, or for some protocols such as QKD by correcting the outcome of the protocol during classical post-processing using the measurement results received from the helpers.

\begin{figure}
     \centering
        \subfigure[]{\includegraphics[width=0.33\textwidth]{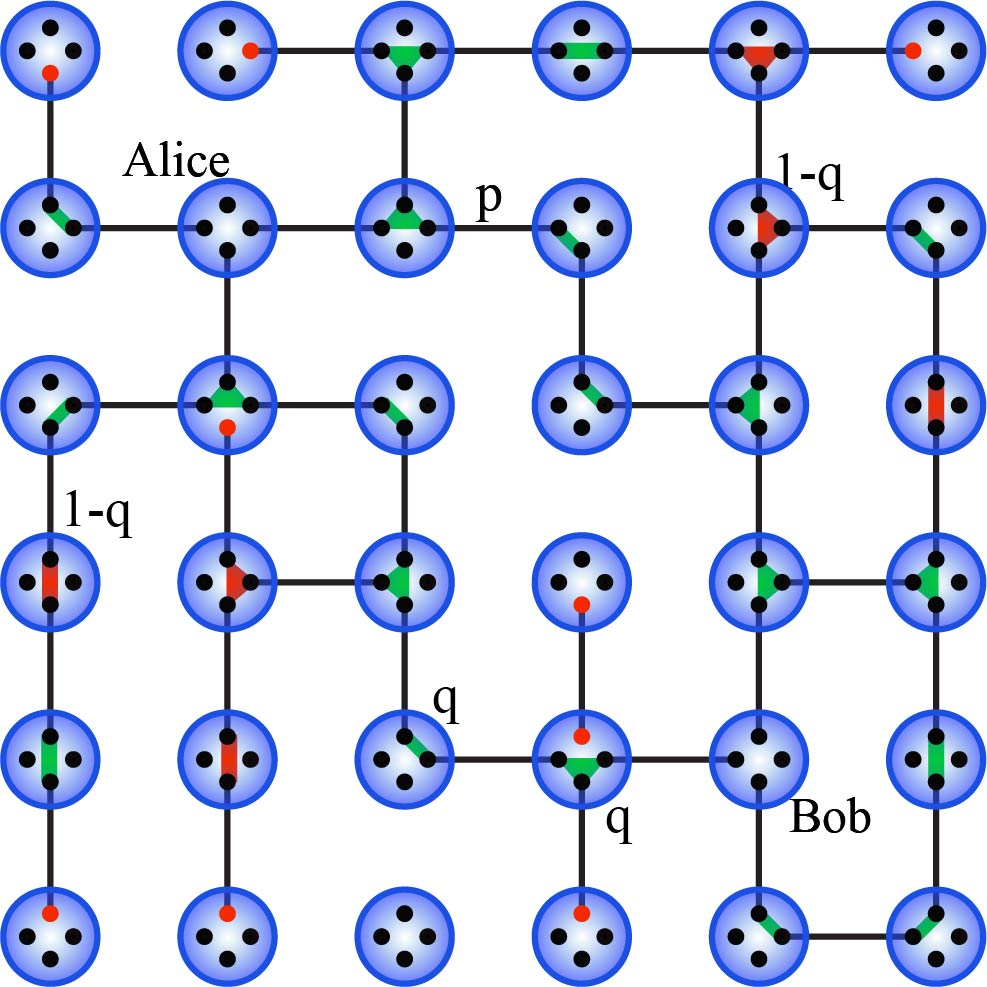}\label{fig:protocol step1}}
     \hfill
        
        \subfigure[]{\includegraphics[width=0.33\textwidth]{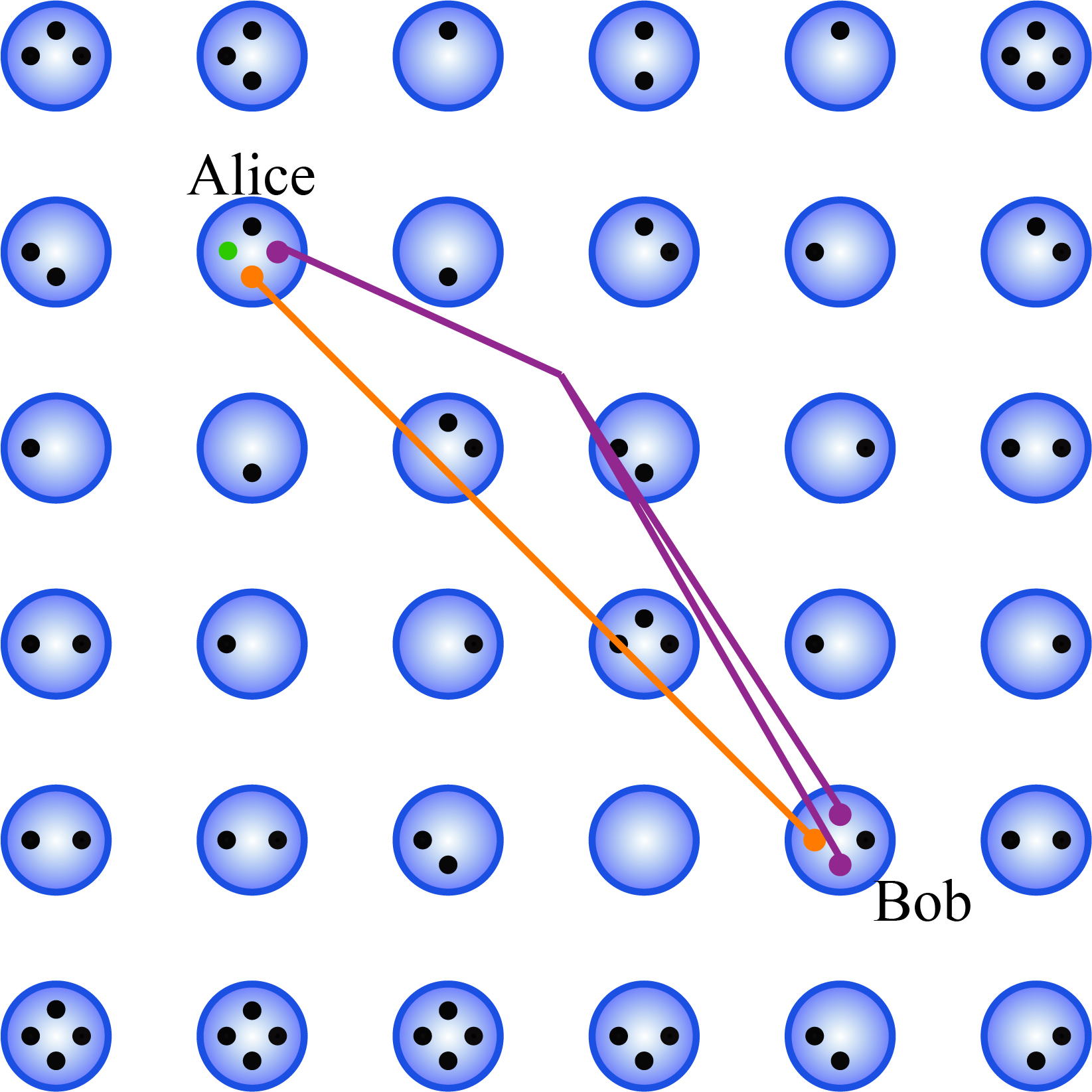}
         \label{fig:protocol step3}}
     \caption{The schematic of the quantum network during various stages of the random 3-GHZ protocol. (a) The quantum network after link generation. The successfully generated links are shown using black solid lines. The green triangles and  rectangles denote the successful fusions. Their red counterparts represent the failed fusion attempts. The quantum memories marked in red perform X-basis measurements on the qubits held in them. (b) The three GHZ states (green, orange, and magenta lines) generated after performing fusions at the repeaters. (b) The $m$-qubit GHZ state shared between Alice and Bob (orange ($m=2$), and magenta ($m=3$) lines).}
        \label{fig:protocol}
\end{figure}

\section{Entanglement rates}
\label{sec:results}
We calculate the shared entanglement generation rate for the square-grid topology of the quantum network under three different fusion rules (Fig. \ref{fig:protocol}) as a function of link and fusion success probabilities $(p,q)$ and the distance between the consumers. We define rate as expected number of $n$-GHZs (including Bell pairs) shared between the consumers per cycle. We can think of the quantum network shown in Fig. \ref{fig:protocol step1} as a graph $G(V,E)$ such that each quantum memory is a vertex $v\in V$, and each link $e\in E$ is created with probability $p$ (a successfully created link Bell pair). Fusion operations are then executed at vertices with at least two neighbors creating a new graph $G'(V',E')$. In this graph, $v\in V'$ is a quantum memory that has undergone a fusion operation. $G'(V',E')$ has additional edges that represent the edges created between quantum memories at the same repeater due to successful fusions between vertices (green triangles or rectangles in Fig. \ref{fig:protocol step1}).  In addition, consumers Alice and Bob have four vertices each. They share an entangled state at the end of the fusion stage, if they belong to the same connected component of graph $G'$. The number of GHZ states shared between Alice and Bob equals the number of disconnected sub-graphs of $G'$ containing at least one vertex each from both Alice and Bob. Hence, the maximum value the rate can take would be 4 $m$-qubit GHZ states/cycle where $m\le 8$ for a square grid network. In the following sections, we compute and compare the shared entanglement generation rates for the different protocols over square-grid and random networks. We refer to the protocol in which repeaters can perform up to $n$-qubit GHZ projections as $n$-GHZ protocol.

\subsection{Perfect repeaters}
We first study the case where repeaters always successfully perform fusions, i.e., $q=1$. In the $n$-GHZ protocol over a certain network topology, calculating the probability that the consumers are a part of the same connected component of $G'(V',E')$ translates to a bond percolation problem on the underlying network topology~\cite{grimmett1999percolation}. The link generation probability $p$ is equivalent to the bond occupation probability in the percolation problem. Percolation is a phase transition phenomenon such that when $p<p_c$ (sub-critical regime), where $p_c$ is a threshold that depends on the lattice geometry, the probability that two randomly chosen sites are connected decays exponentially with distance between the two sites. On the other hand, if $p>p_c$ (super-critical regime), this probability remains constant with the distance.
This result forms the basis of our protocols to achieve distance-independent shared entanglement generation rates. 

\begin{figure}
     \centering
     \subfigure[]{\includegraphics[width=0.5\textwidth]{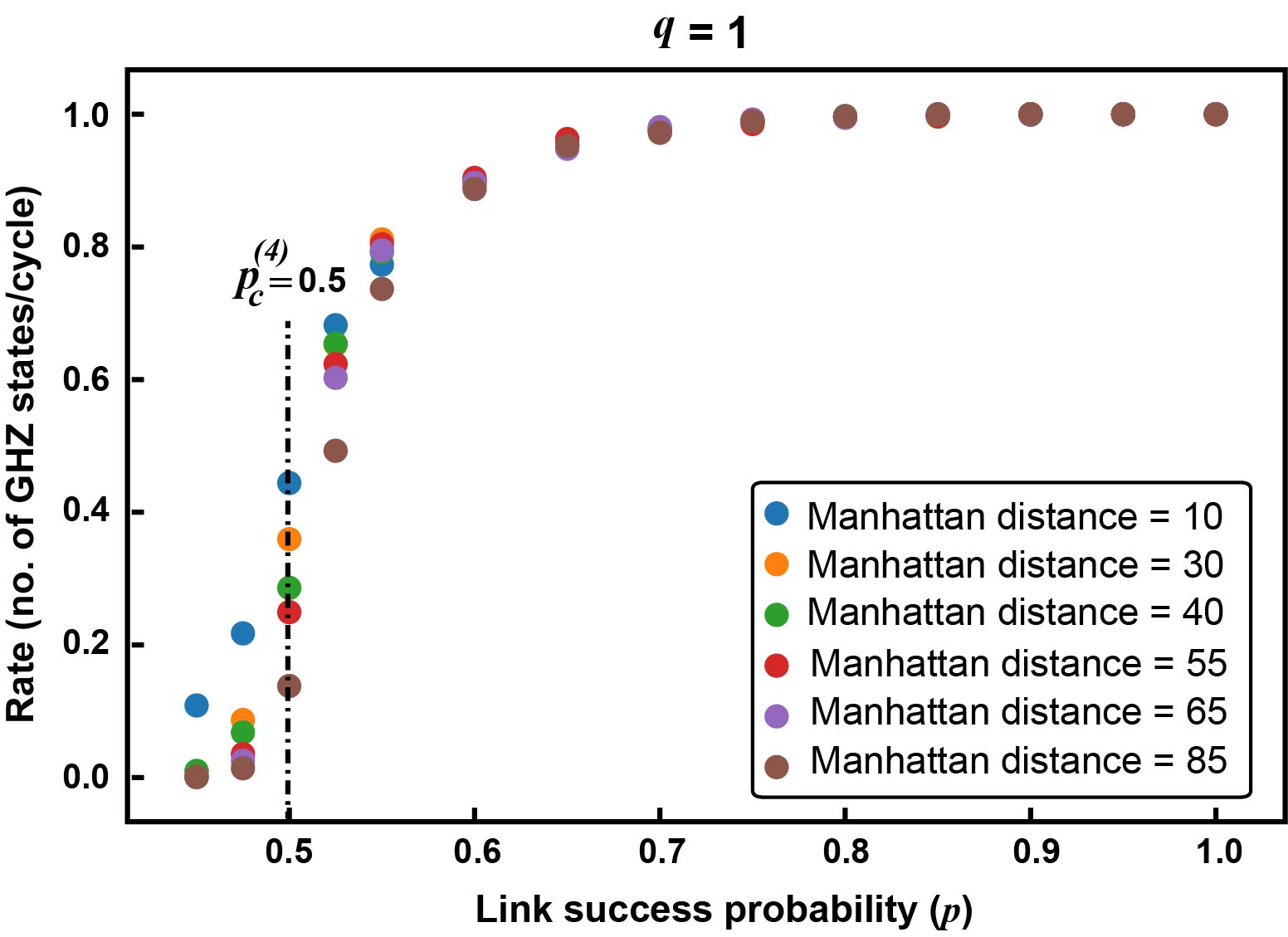}\label{fig:4ghz_rate_p}}
     \hfill
     \subfigure[]{\includegraphics[width=0.5\textwidth]{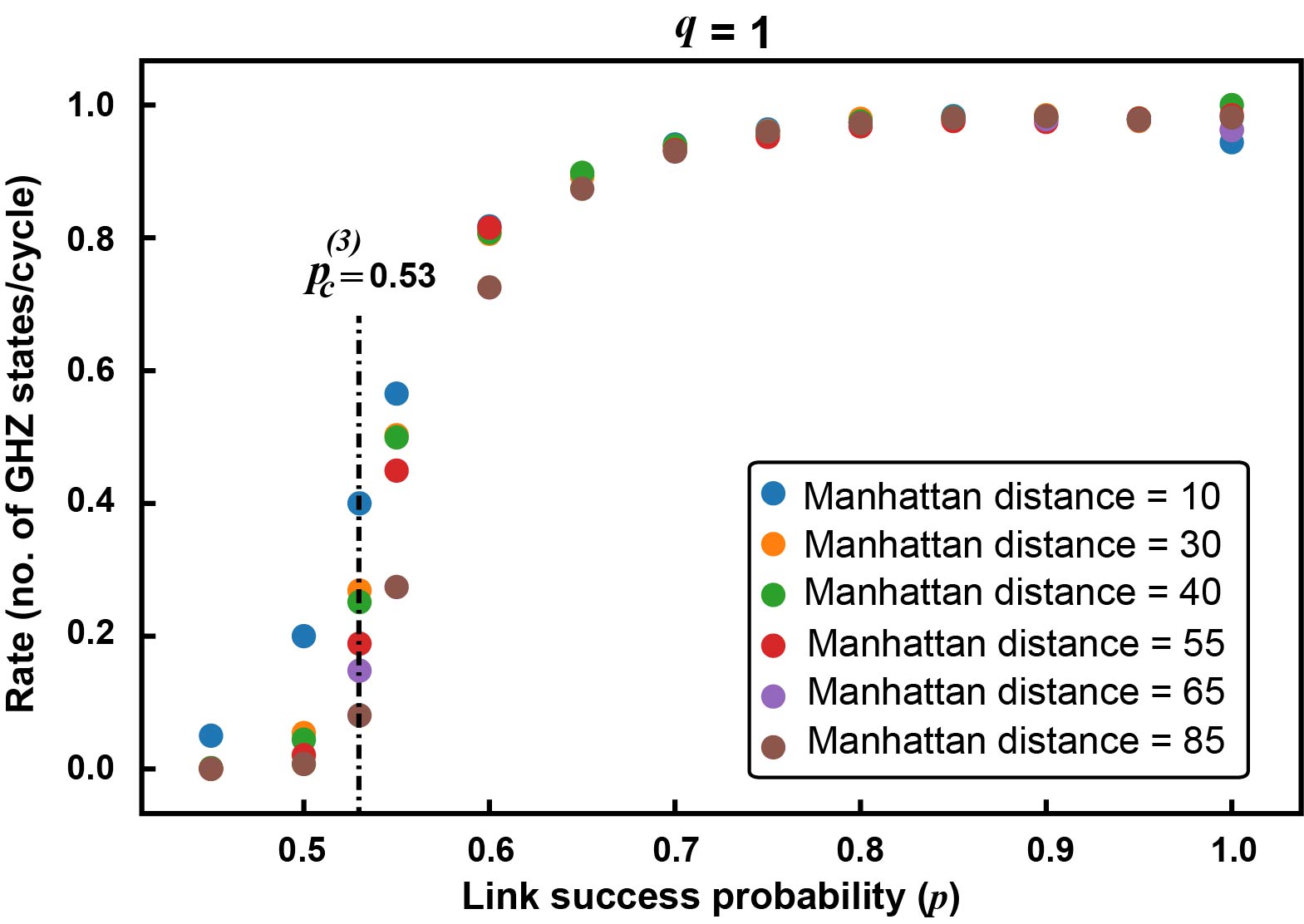}\label{fig:3ghz_rate_p}}
        \caption{Entanglement rate over the square grid network assuming $q=1$ for, (a) the 4-GHZ protocol (b) the random 3-GHZ protocol. We see that (a) above the threshold $p > p_c$, the entanglement rate becomes independent of the distance between communicating parties while it scales with the Manhattan distance when $p<p_c$; and (b) the threshold $p_c$ is higher ($0.53$ versus $0.5$) for the 3-GHZ protocol. The threshold $p_c$ for the 4-GHZ protocol is the standard bond-percolation threshold of the 2D square lattice.}
        \label{fig:perfect repeaters}
\end{figure}

In Fig. \ref{fig:4ghz_rate_p}, we plot the expected number of GHZ states shared by the consumers---at different Manhattan-distance separations---at the end of each cycle for the 4-GHZ protocol, as a function of $p$. As expected, we see that as $p$ goes above the bond percolation threshold of the square grid, $p_c^{(4)} = 0.5$, (a) the rate increases sharply, and (b) the rate becomes independent of the separation between the consumers.

The 3-GHZ protocol described above translates to a different bond percolation problem on the square lattice, where up to $3$ occupied bonds incident at a node can be stuck together to form connected components. For this problem, the bond percolation threshold $p_c^{(3)} \approx 0.53$ (Fig. \ref{fig:3ghz_rate_p}). For both of these fusion rules at the repeaters, when $p>p_c^{(3)}$, the rate doesn't decay exponentially  with the distance between the consumers, but remains constant instead.

\subsection{Imperfect repeaters}
\begin{figure}
     \centering
     \subfigure[]{\includegraphics[scale=0.6]{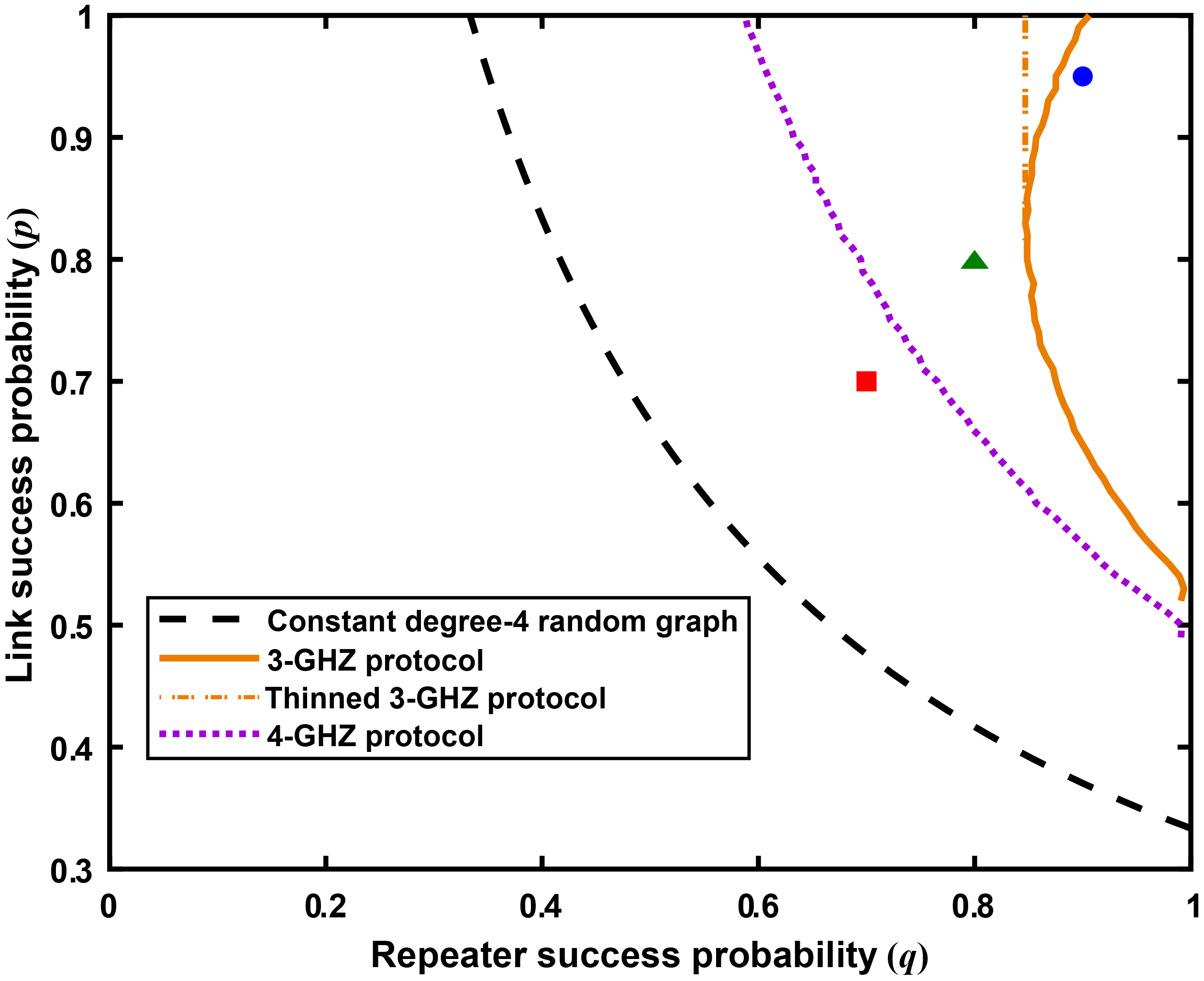}\label{fig:site_bond}}
     \hfill
     \subfigure[]{\includegraphics[width=0.44\textwidth]{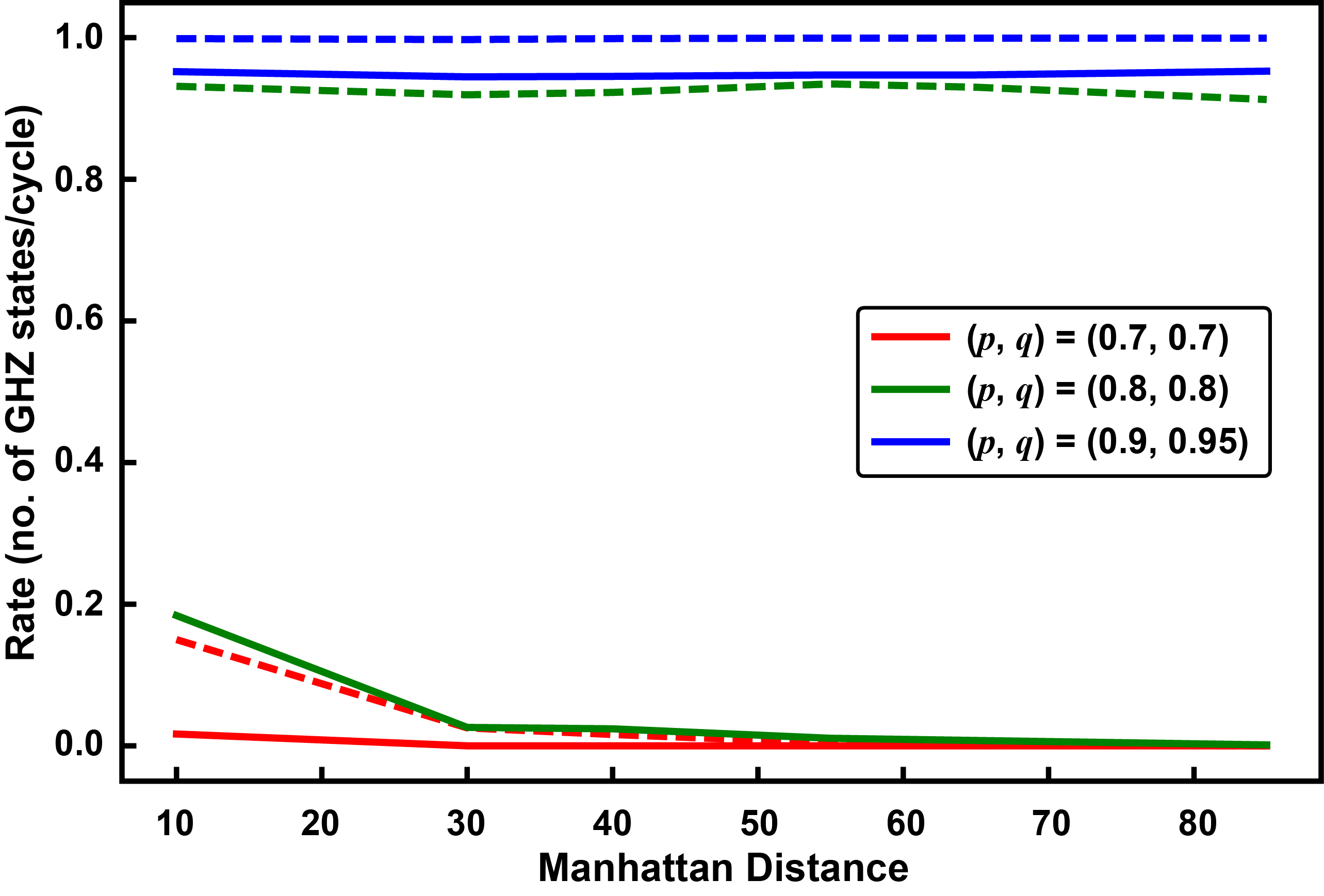} \label{fig:rate_d_p_q}}
        \caption{(a) Site-bond percolation regions for the percolation problems corresponding to the 4-GHZ protocol over square-grid network, 3-GHZ protocol over square-grid and constant degree-4 random graph networks. The curves represent the critical regime of percolation. $p$ and $q$ need to lie above the curves for distance independent rate.  (b) Rate vs distance for points in three different regions of the site-bond curve marked in (a) for the square-grid network. The dashed and solid lines correspond to 4- and random 3-GHZ protocols respectively. }
\end{figure}

Depending on the quantum hardware used at the repeaters, fusion operations can be probabilistic~\cite{Ewert2014-nr}. In this paper, if a repeater fails to perform fusion, it is equivalent to performing X-basis measurements on the qubits involved in the fusion. Calculating the probability that a pair of users end up with shared entanglement when both link generation and fusions are probabilistic, now, becomes a site-bond percolation problem~\cite{Hammersley1961-op} over the underlying network topology lattice (e.g. the square lattice). Site-bond percolation is the generalized version of a percolation problem in which sites and bonds are present with probabilities $q$ and $p$ respectively. The boundary between the super- and sub- critical regimes becomes a curve in the $(p,q)$ plane. For our protocol, the fusion success probability at each repeater translates to the site occupation probability $q$. Here, we assume that all fusion operations succeed with the same probability $q$. We analytically calculate the site-bond region for an $n$-GHZ protocol over a random graph in Section \ref{apx:nGHZ_conf_graph}. Fig. \ref{fig:site_bond} shows the site-bond region for the lattices formed after the fusion step in 3- and 4-GHZ projection protocols on a square-grid network, simulated using the Newman-Ziff method~\cite{Newman2001-vm} and $3$-GHZ protocol on a constant degree-4 random graph network using the analytical formula. The site-bond curve gives the percolation thresholds $(p_c,q_c)$ of the underlying lattice. The probability that the two consumers are connected is distance-independent when $p>p_c$ and $q>q_c$. Thus, the link generation and fusion success probabilities need to lie above the site-bond curve to achieve distance-independent rate. To demonstrate this, we plot the rate as a function of distance for three pairs of $(p,q)$ that lie in three different regions of the site bond curves  of the 4- and 3- GHZ protocols in Fig. \ref{fig:rate_d_p_q}.

\begin{figure}

     \centering
         \includegraphics[scale=0.6]{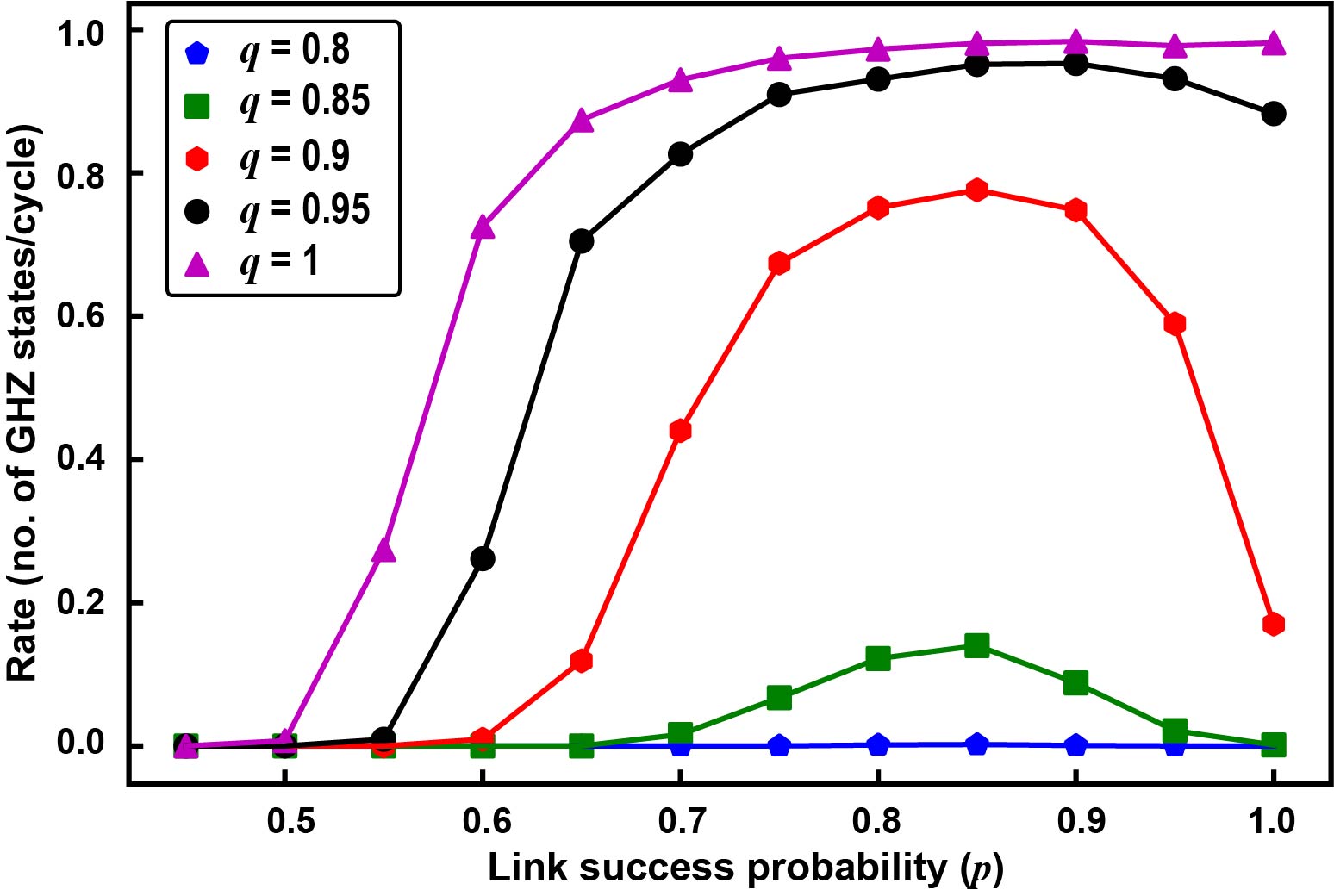}
         \caption{Rate vs link success probability for the 3-GHZ protocol over the square-grid network }
        \label{fig:3ghz_rate_p_q}
        \end{figure}
        
\subsection{Rate upper bound}\label{sec:rateUB}
\begin{figure}
    \centering
    \includegraphics[scale=0.65]{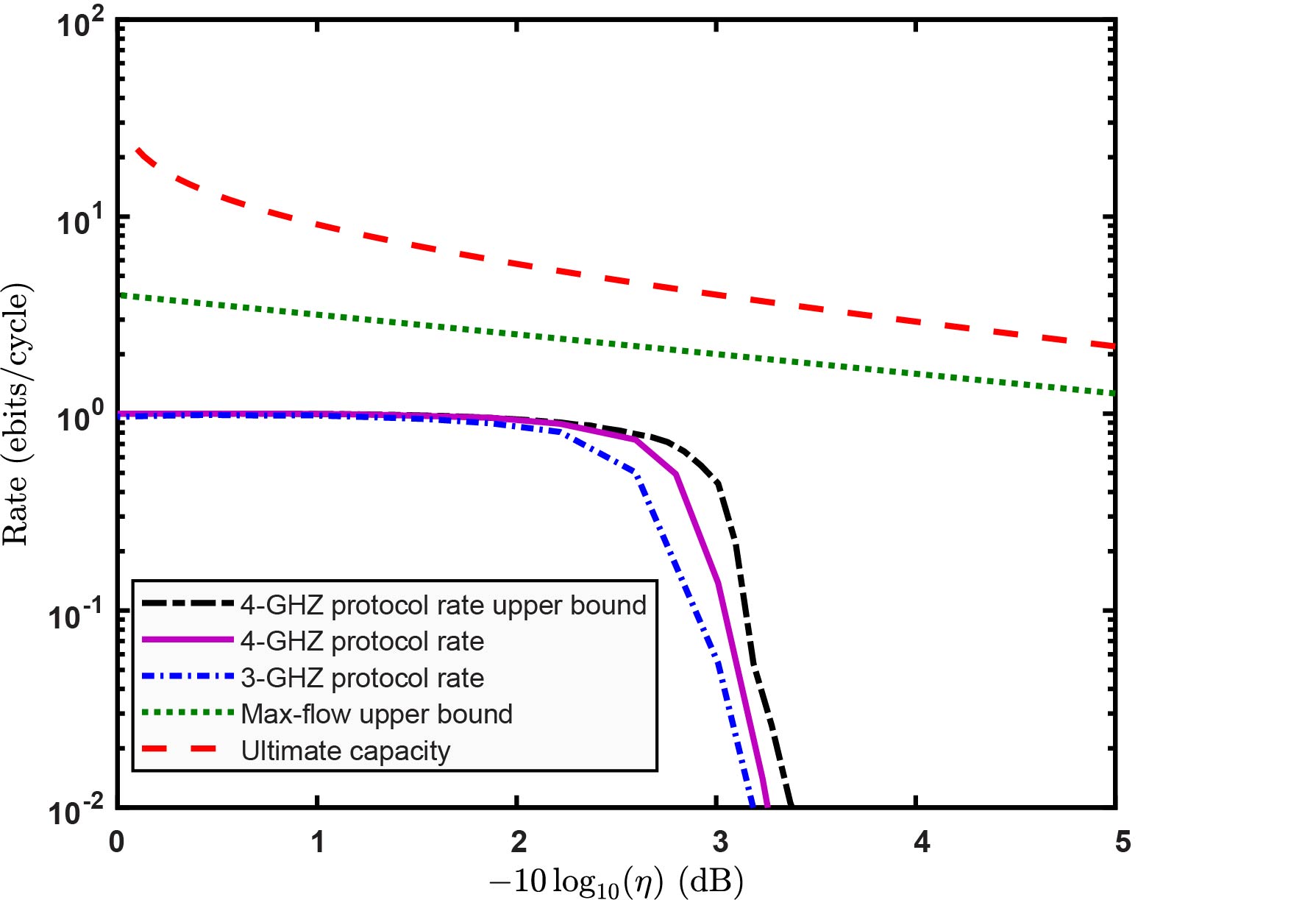}
    \caption{Upper bound to our 4-GHZ protocol’s rate $F(p)^2$, the actual rate attained by our protocol when nodes use $4$-GHZ projections, the actual rate attained by our protocol when nodes use $3$-GHZ projections, upper bound from \cite{pirandola2019end} given by the max-flow, and the ultimate capacity ($-4\log(1-\eta)$) as a function of channel transmissivity $\eta$. Alice and Bob are assumed to one hop away from the diametrically-opposite corner points of a $100$-by-$100$ square grid.
}
    \label{fig:capacity_comparison}
\end{figure}

Ref.~\cite{pirandola2019end} gave the capacity upper bound for a quantum network with repeaters that can perform arbitrary quantum operations, including multi-qubit projective measurements. This work transforms the entanglement routing protocol to a flow problem. A pair of achievability and converse proofs established that the capacity of generating entanglement between two network nodes Alice and Bob is given by the max-flow (or the min-cut) of the underlying network topology, where the capacity of an individual link of transmissivity $\eta$ is $-\log_2(1-\eta)$ ebits per mode~\cite{Pirandola2017-ej}. This ultimate capacity of a square-grid network topology with equal link capacities, translates to $-4\log_2(1-\eta)$ ebits/mode (see red dashed plot in Fig.~\ref{fig:capacity_comparison}). This hence must serve as an upper bound for rates achieved by our protocol. 

The most natural realization of the protocols presented in this paper would be with single-photon dual-rail (e.g., polarization-encoded) qubits. For this qubit encoding, if everything else other than the link loss were ideal (unit-efficiency memory, perfect detectors, deterministic entanglement sources with no multi-pair emissions), the success probability of a link, per mode, $p = \eta$. When $q=1$, the max-flow is given by $4p$ (see green dotted plot in Fig.~\ref{fig:capacity_comparison}), which serves as an upper bound to the rates achieved by our specific protocols. To achieve the high rates at low loss predicted by ~\cite{pirandola2019end}, one must use high-order, e.g., continuous-variable modulation formats.

This upper bound of $4p$ is clearly independent of the distance between the consumers Alice and Bob, which is consistent with our protocol’s having such a property. Ref.~\cite{pirandola2019end}'s result implies that this upper bound would be achievable with perfect quantum memories and perfect BSMs at all nodes. This is simple to see since the link-level entangled qubit pairs can be stored perfectly for a long time. So, after $k$ time steps, each link would have accumulated roughly $kp$ successfully entangled Bell states, at which point, they can be connected into $4kp$ end-to-end Bell states along four edge-disjoint paths connecting Alice and Bob using BSMs, achieving a rate of $4kp/k = 4p$ ebits per cycle.

For our protocol, this upper bound is not achievable because we assume limited memory coherence time (equivalent to the classical communication latency of a link), limited quantum processing available at every node, and act without the advantage of global link-state knowledge. Assuming that the consumers are far apart, when $p < p_c^{(l)}$, for $l = 3$ or $l = 4$ (depending upon which of the two protocols is being used), with high probability, they will not share a connected path. When $p > p_c^{(l)}$, the consumers will with high probability, we part of the giant connected component (GCC), a unique $O(N)$-size connected component where $N$ is the total number of bonds in the underlying network. Hence, in this super-critical regime, at the end of each cycle, with high probability, the consumers share a perfect GHZ state.

Let us say that the probability of a random node belonging to the GCC is $F(p)$. In order for Alice and Bob to share a GHZ state at the end of a cycle, they both have to be a part of the GCC. Hence, the rate achieved by our protocol is upper bounded by $F(p)^2$ (see black dash-dotted plot in Fig.~\ref{fig:capacity_comparison}). Finally, we plot the actual rates achieved by our 4-GHZ and 3-GHZ protocols for comparison, in Fig. \ref{fig:capacity_comparison} (solid magenta and blue dash-dotted respectively). 

\section{Improved $n$-GHZ protocol}
\label{sec:improved 3-GHZ}
We observe a curious turnaround in the site-bond curves for the 3-GHZ and some $n$-GHZ protocols over square-grid and random networks, respectively. For the 3-GHZ protocol, when two neighbouring repeaters have four link successes each and they are limited to doing 3-GHZ projections, while one repeater might sacrifice the shared edge, the other might not as the repeaters don't communicate with each other to decide which links to choose to perform fusion on. It effectively disconnected the two repeaters even though they shared a link. This effect is even more pronounced when $q<1$ because a repeater might choose to fuse with a failed neighbouring repeater instead of a functioning one. This negatively affects the overall connectivity of the lattice. As a result of the adversarial nature of the rule, in high $p$ regime, the rate starts decaying with $p$ when $q<1$ (Fig. \ref{fig:3ghz_rate_p_q}), which shows that the rule is sub-optimal. The site-bond region for this 3-GHZ fusion rule clearly depicts this behaviour in Fig. \ref{fig:site_bond}. Similar arguments can be made to explain the turnaround for the $n$-GHZ protocol for the random network. In the following sections, we discuss three strategies to improve the turnaround.
\subsection{Thinning the network}
Let $p^*$ be the link generation probability at which the turnaround occurs. The adversarial behaviour of the protocol is observed only beyond $p^*$. We can get rid of the turnaround by randomly removing links in the high $p$ regime. We modify the protocol such that when $p>p^*$, each link is deleted with probability $(p-p^*)/p$. This makes the effective link generation probability $p^*$ when $p>p^*$ as shown in FIG. \ref{fig:site_bond}.
\subsection{The Brickwork network}
\label{sec:brickwork}
\begin{figure}
     \centering
     \subfigure[]{\includegraphics[width=0.33\textwidth]{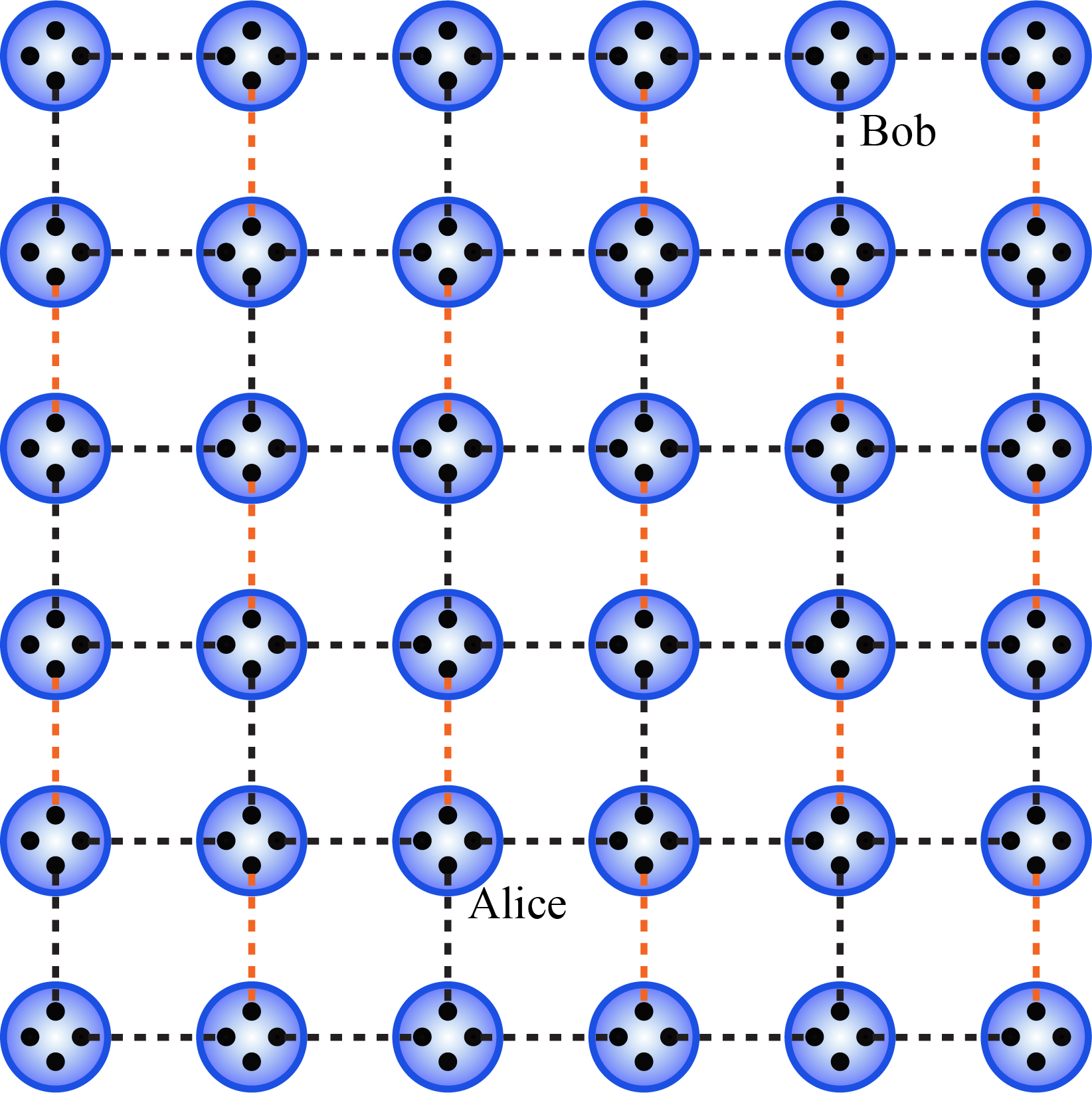}
         \label{fig:brickwork_protocol0}}
    \hfill
     \subfigure[]{\includegraphics[width=0.33\textwidth]{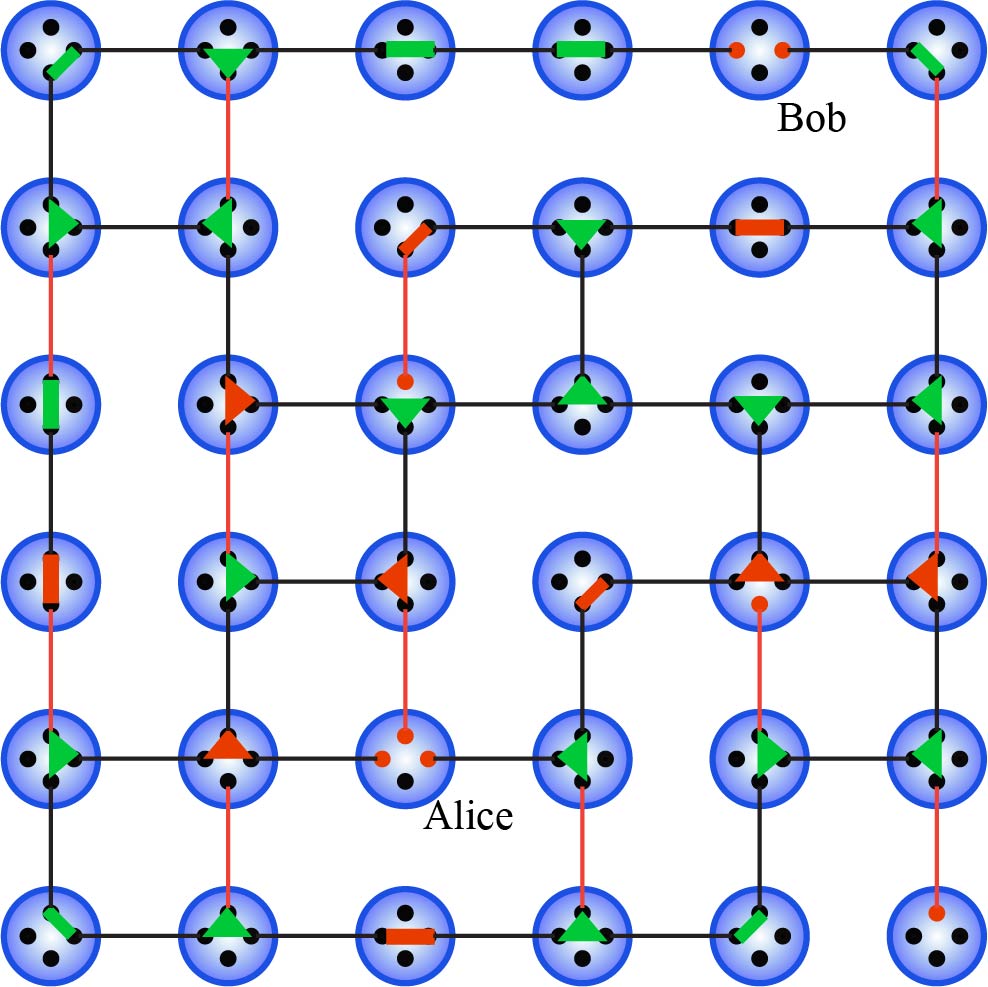}
         \label{fig:brickwork_protocol}}
     \hfill
     \subfigure[]{\includegraphics[width=0.45\textwidth]{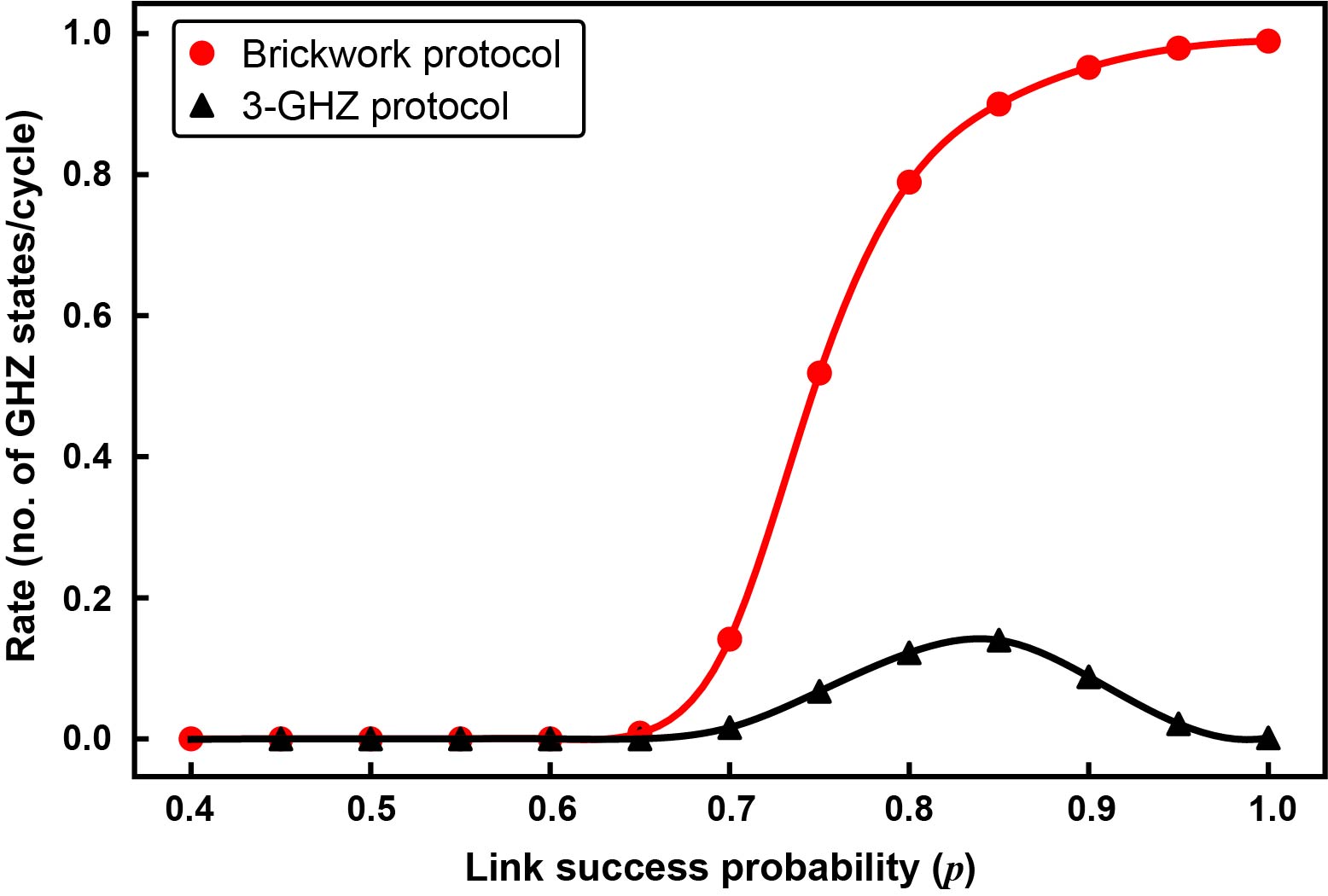}
         \label{fig:brickwork_rate_p_q}}
     \caption{The brickwork protocol over the square-grid network - (a) The dotted lines show the red and black link generation attempt. The dotted black lines form a brickwork lattice. (b) Red links are used for fusion only if there are less than 3 black links present. (c) Comparison between the 3-GHZ protocol and the brickwork protocol for $q=0.85$ and Manhattan distance = 85 units. The rate for the brickwork protocol does not degrade when $p$ is high and $q<1$ but stays constant irrespective of distance.}
        \label{fig:brickwork}
\end{figure}

\begin{figure}
    \centering
    \includegraphics[scale=0.65]{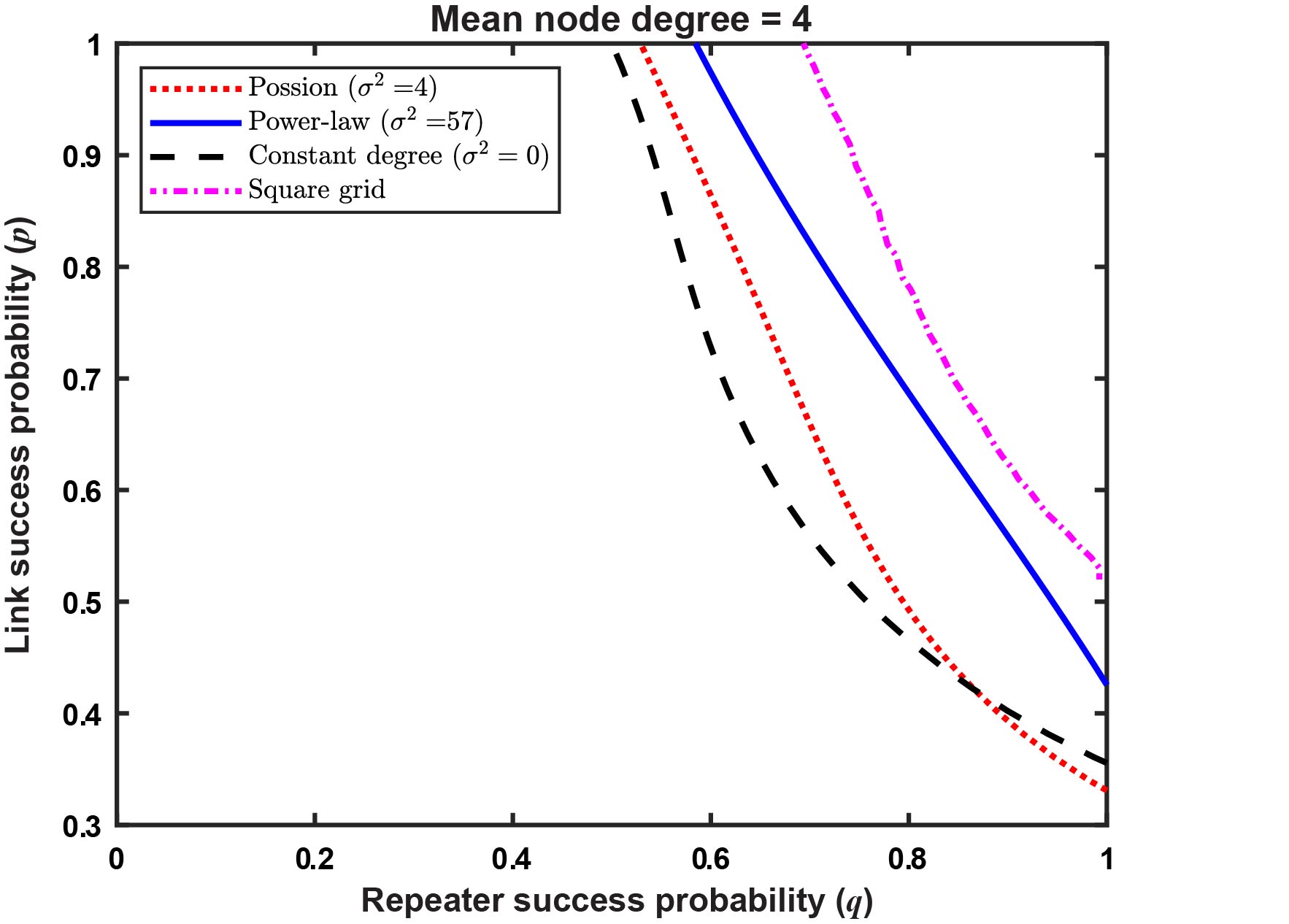}
    \caption{The site-bond region for the 3-GHZ brickwork protocol for the square-grid, Poisson-degree distributed random graph with mean degree 4, and constant degree-4 random graph network topologies}
    \label{fig:compare_configuration_brickwork}
\end{figure}

The random selection of the links to fuse degrades the rate when repeaters can fail. To overcome this issue, we propose a deterministic link selection rule that doesn't let neighbouring repeaters make conflicting fusions. Consider the square-grid topology of the quantum repeaters. This network has two types of links - red and black. Both red and black links have the same success probability $p$. Links are arranged such that the black links form a brickwork lattice. Each repeater has a maximum three black links and one red link. In the fusion step of the protocol, a repeater uses the red link only if it has two or fewer black links as shown in Fig. \ref{fig:brickwork_protocol}. This protocol is equivalent to percolation over a brickwork lattice with an extra optional bond at each site. Hence, we observe in FIG. \ref{fig:compare_configuration_brickwork} that the repeater success probability threshold is equal to the site percolation threshold of the brickwork lattice. And the link success probability is higher than the bond percolation threshold of the brickwork lattice due to the additional bond. This fixed selection rule gets rid of the adversarial nature of the previous protocol without having neighbouring repeaters communicate with each other. Fig. \ref{fig:brickwork_rate_p_q} shows the rate vs. link success probability ($p$) curve doesn't decay when the repeaters fail to perform fusions ($q<1$).

The brickwork model can be adapted for random graphs as well by dividing the edges into two categories - black and red. The lattice formed by the black edges is not a brickwork lattice in this case. For the $n$-GHZ protocol over a random network to make the protocol partially deterministic, each node can have maximum $n$ black edges and the rest are red edges. If the total number of edges at a node is less that $n$, all of them are black. Each repeater (node) uses the red links for fusion only if it has less than $n$ black links. We compare the site-bond regions for the 3-GHZ brickwork protocol for various network topology with mean degree $\approx 4$. We observe that configuration graphs do better than the square-grid as they offer long-range connectivity. We notice that although this strategy improves the site-bond region, it doesn't remove the turnaround for all combinations of network topologies and $n$ as shown in Fig. \ref{fig:configuration_brickwork}. The analytical expression for the site-bond region of this brickwork-like model for random graphs is derived in Section \ref{apx:brickwork_conf_graph}.
\subsection{Dividing the network}
\label{subsec:dividing the network}
As discussed earlier, the entanglement generation rate is proportional to the number of disconnected sub-graphs of the graph generated after fusion ($G'(V,E)$) that are shared between the consumers. In the high $(p,q)$ regime, for the square-grid network, due to 3-/4-qubit projections, as the overall connectivity of $G'(V,E)$ improves, its disconnected sub-graphs start merging together. Hence, this framework fails to achieve the maximum rate possible for the underlying network topology, (4 GHZ states/cycle in our case). When $p$ and $q$ both equal one, we end up with only one GHZ state shared between the consumers. This issue can be overcome by dividing the network into four disconnected sub-graphs such that exactly one quantum memory from each consumer resides in one sub-graph. The sub-graphs are never allowed to merge into each other by permanently erasing the edges joining them.



\section{Quantum key distribution}
\label{sec:QKD}
In this section, we briefly discuss a Quantum Key generation protocol to share a secret key between a pair of users using GHZ states. This protocol is an extension of the BBM'92 quantum key distribution protocol~\cite{Bennett1992-bm}. The protocol consists of the following steps -  
\begin{itemize}
    \item \textit{Step 1:} Alice and Bob start with multiple $m+l\ge2$ qubit GHZ states such that Alice and Bob have $m$- and $l$ qubits of the GHZ state. 
    $$\ket{\psi_{AB}}=\frac{\ket{0}^{\otimes m}_{A}\ket{0}^{\otimes l}_{B}+\ket{1}^{\otimes m}_{A}\ket{1}^{\otimes l}_{B}}{\sqrt{2}}$$ 
    Here, $m$ and $l$ can vary across the collection of shared GHZ states Alice and Bob possess.
    \item \textit{Step 2: }They independently and randomly choose between the computational basis (0/1-basis) and the Hadamard basis (+/- -basis) for measurement. Each user measures all their qubits of the GHZ state using (their) randomly-chosen measurement. Alice and Bob get  $m-$ and $l$ - bit results, respectively, after performing the measurements.
    \item \textit{Step 3: }They use a classical channel to inform each other the basis they have used to measure their respective qubits. The measurements instances where Alice and Bob used the same basis are used for key generation. This step is similar to the BBM’92 protocol.
    \item \textit{Step 4: }If both of them used the computational basis in a given round of the protocol, they get bit string of either all 0's or all 1's. In this case, that bit becomes the key. When Alice and Bob both use the Hadamard basis, they get measurement outcome bits strings $a_1a_2\dots a_m$ and $b_1b_2\dots b_l$, respectively, such that $(a_1+a_2+\dots+a_m)\mod{2}=(b_1+b_2+\dots+b_l)\mod{2}$. Here, the key would be the parity of their respective bit strings.
\end{itemize}  
We leave the security proof for this protocol as an open question. But we believe that it can be done as an extension of the security proof for BBM'92. 

\begin{table}[!htb]
    
    \begin{minipage}{.5\linewidth}
      \centering
        \begin{tabular}{!{\vrule width 1pt}c !{\vrule width 1pt} c !{\vrule width 1pt} c !{\vrule width 1pt}} 
 \noalign{\hrule height 1pt}
 \multicolumn{3}{!{\vrule width 1pt}c!{\vrule width 1pt}}{Alice} \\ \noalign{\hrule height 1pt}
 Basis & Output bits & Key\\ [0.5ex] 
 \noalign{\hrule height 1pt}
 +/- & 1010 & 0 \\ 
 \hline
 0/1 & 0 & 0 \\
 \hline
 +/- & 1101 & -\\
 \hline
 +/- & 100 & 1\\
 \hline
 0/1 & 1111 &-\\ 
 \hline
 0/1 & 1 &1\\ 
 \hline
 0/1 & 00 &-\\ 
 \noalign{\hrule height 1pt}
\end{tabular}
    \end{minipage}%
    \begin{minipage}{.5\linewidth}
      \centering
        \begin{tabular}{!{\vrule width 1pt}c !{\vrule width 1pt} c !{\vrule width 1pt} c !{\vrule width 1pt}} 
 \noalign{\hrule height 1pt}
 \multicolumn{3}{!{\vrule width 1pt}c!{\vrule width 1pt}}{Bob} \\ \noalign{\hrule height 1pt}
 Basis & Output bits & Key\\ [0.5ex] 
 \noalign{\hrule height 1pt}
 +/- & 0 & 0 \\ 
 \hline
 0/1 & 0 & 0 \\
 \hline
 0/1 & 11 & -\\
 \hline
 +/- & 010 & 1\\
 \hline
 +/- & 01 &-\\ 
 \hline
 0/1 & 111 &1\\ 
 \hline
 +/- & 110 &-\\ 
 \noalign{\hrule height 1pt}
\end{tabular}
    \end{minipage} 
    \caption{Quantum key generation from shared GHZ-states using the protocol described in \ref{sec:QKD}. When Alice and Bob both use the 0/1 basis, the secret key bit is the bit repeated in the output bit-string. When both of them use the +/- basis, the secret key bit is the parity of their respective output bit-strings.}
\end{table}


\section{Conclusion}
\label{sec:conclusion}
We have designed a quantum-network-based entanglement generation protocol, which affords a rate that is independent of the distance between the users. The protocol only uses local link state information, and has the aforesaid property of distance-independent entanglement rate in a certain region of the link-level entanglement success probability $p$ (which is proportional to the link's optical transmissivity, and hence range) and an individual repeater's success probability $q$ (in performing an $n$-GHZ projective measurement). This $(p,q)$ region that achieves distance-independent rate is the site-bond region of a modified mixed percolation problem, defined on the underlying network such that the bond and site occupation probabilities are given by the link generation and repeater success probabilities, respectively. Our protocol requires only certain local Clifford operations, Pauli measurements, and classical communications. We perform multi-qubit projections at each node of the 2D network making it a multi-path routing protocol. It outperforms the multi-path routing protocol that only uses Bell state measurements (BSMs) \cite{pant2019routing}. All BSM based entanglement protocols exhibit rates that decay with distance even those that use non-local-link state knowledge. To study our protocol for complex quantum networks, we analytically derived the site-bond region for a configuration-model random network with an arbitrary node degree distribution. This shows an excellent match with the numerically-evaluated site-bond region of our modified mixed percolation problem using the Newman-Ziff algorithm. We also discussed a two-party quantum key distribution protocol that can be implemented using the shared entangled state obtained from the entanglement generation protocol.

A few other questions that can be solved as an extension of this protocol are - (1) generating shared entanglement between multiple consumer pairs simultaneously (2) The repeater failure model we have assumed here is very simple. One can study more realistic models repeater failure due to unsuccessful fusions, photon loss, etc.

\section{Methods}
\subsection{Site-bond region for $n$-GHZ protocol over configuration graph}
\label{apx:nGHZ_conf_graph}
\begin{figure}
    \centering
    \includegraphics[scale = 0.8] {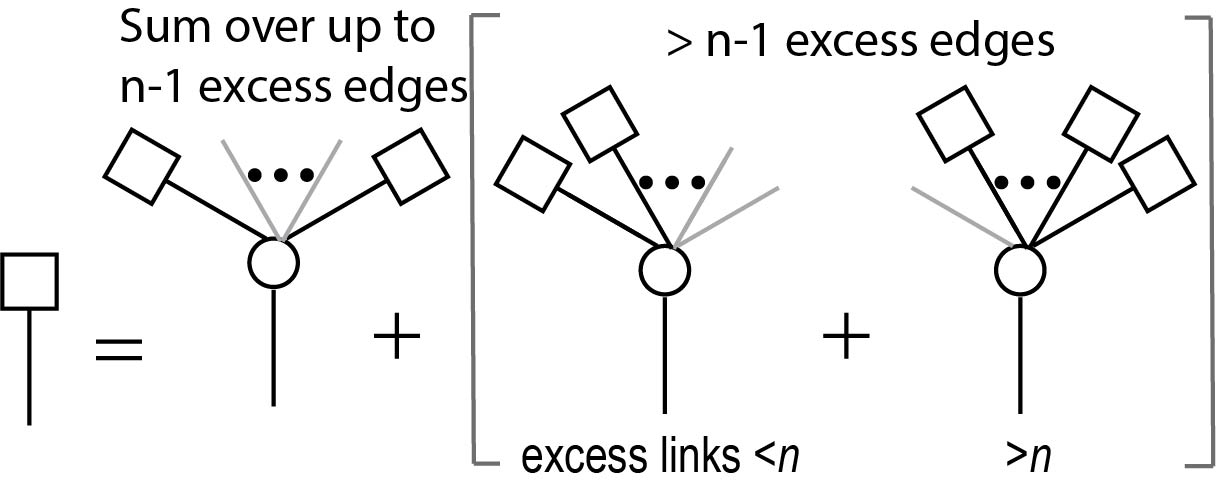}
    \caption{Schematic representation of the sum rule for the connected
component of vertices reached by following a randomly chosen
link.  }
    \label{fig:configuration_graph_model}
\end{figure}
Consider a configuration graph with node degree distribution given by the generating function
\begin{equation}
\label{eq:g0}
G_0(x)= \sum_{d=0}^\infty p_dx^d
\end{equation}
where $p_d$ is the probability that a randomly chosen node has degree $d$. The average node degree is $z=\sum_{d=0}^\infty dp_d$. The generating function for the excess degree distribution is
\begin{equation}
    \label{eq:g1}
    \begin{split}
G_1(x) &= \frac{G_0'(x)}{z}=\frac{\sum_{d=0}^\infty dp_dx^{d-1}}{z} \\
 & = \sum_{d=0}^\infty e_dx^d
\end{split}
\end{equation}

In the percolation problem for the $n$-GHZ protocol, we are allowed to perform up to $n$-qubit GHZ projection at each node (repeater). We start with a random graph with node degree distribution given by (\ref{eq:g0}). In this graph, each edge is occupied with probability $p$, the link generation probability. We call the edges that are occupied ``links". In this section, we derive the site-bond region for a configuration-graph random network by generalizing the formalism in \cite{Newman2001-fz}.

Let $H_1(x)$ be the generating function for the distribution of the size of the component that is reached by choosing a random link and counting all of the nodes that can be reached through one of its end points. Fig. \ref{fig:configuration_graph_model} shows the schematic of the sum rule for $H_1(x)$, the connected component (square) reached by following a randomly chosen link (black lines). We denote nodes by circles and unoccupied edges by grey lines. The distribution of connected component consists of a node at the end of the link we started with and clusters attached (squares) to the node via links (if any). We refer to the node reached by following the link as vertex. The size of the component is zero, if the fusion fails at the vertex with probability $(1-q)$. Excess edges are the edges of a node other than the one used to reach the node. This definition can be extended to excess links as well.
Because of the $n$-GHZ fusion rule, a link always leads to a connected component as long as the number of excess edges at its vertex ($k$) is less than or equal to $n-1$ and the fusion at the vertex is successful. If $k>n-1$, it leads to two possible scenarios - (1) when the excess links at the vertex $l\leq n-1$. In this case, the link connects to a connected component if the fusion succeeds. (2) when $l>n-1$, the size of the component is non-zero if the vertex chooses the link we started with as one of the links for fusion. This happens with probability $n/(l+1)$. When the link is excluded from the fusion, the size of the connected component becomes zero. As we are following a link and not a node, we are interested in the distribution of excess links at the vertex. The probability that a node with $k$ excess edges has $l$ excess links, given each edge is occupied with probability p is -
\begin{equation}
    \label{eq:link_binomial}
    P(l|k)=\binom{k}{l}p^l(1-p)^{k-l}
\end{equation}   
Assuming the fusion success probability is $q$, we write down the sum rule for $H_1(x)$ - 
\begin{equation}
    \label{eq:h1}
    \begin{split}
         H_1(x) &=1-q+qx\sum_{k=0}^{n-1}e_k\sum_{l=0}^kP(l|k)[H_1(x)]^l\\& +qx\sum_{k=n}^{\infty}e_k\Bigg[\sum_{l=0}^{n-1}P(l|k)[H_1(x)]^l\\
         &+\sum_{l=n}^kP(l|k)\frac{n}{l+1}[H_1(x)]^{n-1}\Bigg]\\
         &+q\sum_{k=n}^{\infty}e_k\sum_{l=n}^kP(l|k)\frac{l+1-n}{l+1}
    \end{split}
\end{equation}

The generating function for the distribution of the size of the component that a random node belongs is 
\begin{equation}
    \label{eq:h0}
    H_0(x) = 1-q+qx\sum_{k=0}^{\infty}p_k\sum_{l=0}^{k}P(l|k)[H_1(x)]^l
\end{equation}
and the mean component size is 
\begin{equation}
    \label{eq:mean_cluster_size}
    \langle s \rangle = H_0'(1)=q\sum_{k=0}^{\infty}p_k\sum_{l=0}^{k}P(l|k)(1+lH'_1(1))
\end{equation}
$H_1'(1)$ diverges when 
\begin{equation}
    \label{eq:site-bond curve}
    \begin{split}
        q(p) &\ge \sum_{k=0}^{n-1}e_k\sum_{l=0}^klP(l|k) +\sum_{k=n}^{\infty}e_k\Bigg[\sum_{l=0}^{n-1}lP(l|k)\\
         &+\sum_{l=n}^kP(l|k)\frac{n(n-1)}{l+1}\Bigg]
    \end{split}
\end{equation}
This marks the phase transition for percolation and (\ref{eq:site-bond curve}) gives the site-bond curve.
\begin{figure}[h]
    \centering
    \includegraphics[scale=0.65]{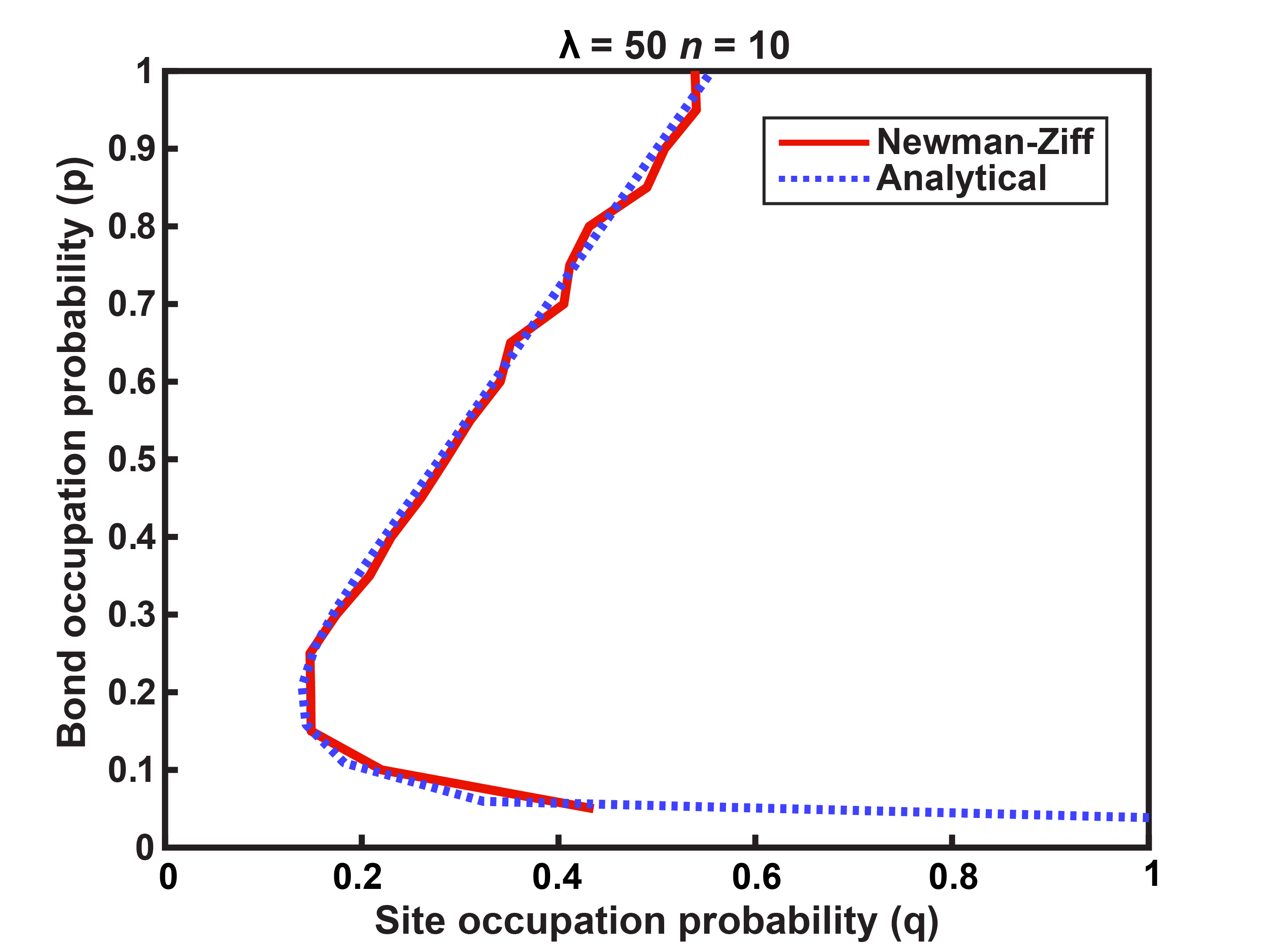}
    \caption{Analytically calculated (using (\ref{eq:site-bond curve})) and simulated site-bond region for the $10$-GHZ protocol over a configuration graph network with Poisson degree distribution with mean $\lambda=50$.}
    \label{fig:site_bond_configuration_graph}
\end{figure}

\subsection{Brickwork-like model for configuration graph}
\label{apx:brickwork_conf_graph}
In the site-bond curve for the $n$-GHZ protocol over a configuration graph network, after a certain value of $p$, the turnaround point, $q$ starts increasing with $p$ as shown in Fig. \ref{fig:site_bond_configuration_graph}. This happens due to the adversarial nature of the protocol explained in \ref{sec:improved 3-GHZ}. In this section, we calculate the site-bond region for the brickwork-like strategy for configuration graphs to improve the entanglement generation rate beyond the turnaround point.

For the $n$-GHZ protocol over a configuration graph network whose degree distribution is given by (\ref{eq:g0}), to make the protocol deterministic, we divide the edges into two categories - black and red. Each node can have maximum $n$ black edges and the rest are red edges. If the total number of edges at a node is less that $n$, all of them are black. Each repeater (node) uses the red links for fusion only if it has less than $n$ black links. Let $H_{11}(x), H_{12}(x)$ be the distribution of the sizes of components that are reached by following black and red links, respectively. Let $l_1$ and $l_2$ be respectively the number of black and red excess links at a node such that $l=l_1+l_2$. 
\begin{equation}
    \label{eq:h11}
    \begin{split}
        H_{11}(x)&=1-q+qx\sum_{k=0}^{n-1}e_k\sum_{l=0}^kP(l|k)[H_{11}(x)]^{l}
        \\&+qx\sum_{k=n}^{\infty}e_k\sum_{l_1=0}^{n-1}\Bigg[\sum_{l_2=0}^{n-1-l_1}p^{l_1+l_2}(1-p)^{k-l_1-l_2}\binom{n-1}{l_1}\\&\times\binom{k-n+1}{l_2}[H_{11}(x)]^{l_1}[H_{12}(x)]^{l_2}
        \\&+\sum_{l_2=n-l_1}^{k-n+1}p^{l_1+l_2}(1-p)^{k-l_1-l_2}\binom{n-1}{l_1}\binom{k-n+1}{l_2}
        \\&\times [H_{11}(x)]^{l_1}[H_{12}(x)]^{n-1-l_1}\Bigg]
    \end{split}
\end{equation}
\begin{equation}
    \label{eq:h12}
    \begin{split}
       H_{12}(x)&=1-q+       qx\sum_{k=n}^{\infty}e_k\sum_{l_1=0}^{n-1}\Bigg[\sum_{l_2=0}^{n-1-l_1}p^{l_1+l_2}(1-p)^{k-l_1-l_2}
       \\&\times\binom{n}{l_1}\binom{k-n}{l_2}[H_{11}(x)]^{l_1}[H_{12}(x)]^{l_2}
        \\&+\sum_{l_2=n-l_1}^{k-n}p^{l_1+l_2}(1-p)^{k-l_1-l_2}
        \binom{n}{l_1}\binom{k-n}{l_2}[H_{11}(x)]^{l_1}\\&\times[H_{12}(x)]^{n-1-l_1}\frac{n-l_1}{l_2+1}\Bigg]+q\sum_{k=n}^{\infty}e_k\sum_{l_1=0}^{n}\sum_{l_2=n-l_1}^{k-n}
        \\&p^{l_1+l_2}(1-p)^{k-l_1-l_2}\binom{n}{l_1}\binom{k-n}{l_2}\frac{l_1+l_2-n}{l_2+1}
    \end{split}
\end{equation}
The distribution of sizes of components to which a randomly chosen node belongs is given by - 
\begin{equation}
    \label{eq:h0_brickwork}
    \begin{split}
        H_0(x)&=1-q+qx\sum_{k=0}^{n}p_k\sum_{l=0}^{k}P(l|k)[H_{11}(x)]^{l}
        \\&+qx\sum_{k=n+1}^{\infty}p_k\sum_{l_1=0}^{n}\Bigg[\sum_{l_2=0}^{n-l_1}p^{l_1+l_2}(1-p)^{k-l_1-l_2}\binom{n}{l_1}
        \\&\times\binom{k-n}{l_2}[H_{11}(x)]^{l_1}[H_{12}(x)]^{l_2}+ \sum_{l_2=n-l_1}^{k-n}p^{l_1+l_2}\\&\times(1-p)^{k-l_1-l_2}
        \binom{n}{l_1}
        \binom{k-n}{l_2}[H_{11}(x)]^{l_1}[H_{12}(x)]^{n-l_1}\Bigg]
    \end{split}
\end{equation}
The average cluster size $\langle s \rangle$ diverges when the giant component appears. 
\begin{equation}
    \label{eq:h'0_brickwork}
    \begin{split}
        \langle s\rangle&=H'_0(1)=q \sum_{k=0}^{n}p_k\sum_{l=0}^{k}P(l|k)\big(1 +lH'_{11}(1)\big)
        \\&+q\sum_{k=n+1}^{\infty}p_k\sum_{l_1=0}^{n}\Bigg[\sum_{l_2=0}^{n-l_1}p^{l_1+l_2}(1-p)^{k-l_1-l_2}\binom{n}{l_1}
        \\&\times\binom{k-n}{l_2}\big(1+l_1H'_{11}(1)+l_2H'_{12}(1)\big)+ \\&\sum_{l_2=n-l_1}^{k-n}p^{l_1+l_2}(1-p)^{k-l_1-l_2}
        \\&\binom{n}{l_1}
        \binom{k-n}{l_2}\big(1+l_1H'_{11}(1)+(n-l_1)H'_{12}(1)\big)\Bigg]
    \end{split}
\end{equation}
\begin{equation}
\label{eq:h11'_brick}
    \begin{split}
        H'_{11}(1)&=q\sum_{k=0}^{n-1}e_k\sum_{l=0}^kP(l|k)\big(1+l_1H'_{11}(1)\big)+q\sum_{k=n}^{\infty}e_k
        \\&\sum_{l_1=0}^{n-1}\Bigg[\sum_{l_2=0}^{n-1-l_1}p^{l_1+l_2}(1-p)^{k-l_1-l_2}\binom{n-1}{l_1}\\&\times\binom{k-n+1}{l_2}\big(1+l_1H'_{11}(1)+l_2H'_{12}(1)\big)
        \\&+\sum_{l_2=n-l_1}^{k-n+1}p^{l_1+l_2}(1-p)^{k-l_1-l_2}\binom{n-1}{l_1}\binom{k-n+1}{l_2}
        \\&\times \big(1+l_1H'_{11}(1)+(n-1-l_1)H'_{12}(1)\big)\Bigg]
    \end{split}
\end{equation}
\begin{equation}
    \label{eq:h12'_brick}
    \begin{split}
       H'_{12}(1)&=      q\sum_{k=n}^{\infty}e_k\sum_{l_1=0}^{n-1}\Bigg[\sum_{l_2=0}^{n-1-l_1}p^{l_1+l_2}(1-p)^{k-l_1-l_2}
       \\&\times\binom{n}{l_1}\binom{k-n}{l_2}\big(1+l_1H'_{11}(1)+l_2H'_{12}(1)\big)
        \\&+\sum_{l_2=n-l_1}^{k-n}p^{l_1+l_2}(1-p)^{k-l_1-l_2}
        \binom{n}{l_1}\binom{k-n}{l_2}\\&\times\frac{n-l_1}{l_2+1}\big(1+l_1H'_{11}(1)+(n-1-l_1)H'_{12}(1)\big)\Bigg]
    \end{split}
\end{equation}
Equations (\ref{eq:h11'_brick}) and (\ref{eq:h12'_brick}) form linear system equations in $H'_{11}(1)$ and $H'_{12}(1)$ and can be re-written as - 
\begin{eqnarray}
        H'_{11}(1)&=&qS_{11}H'_{11}(1)+qS_{12}H'_{12}(1)+C_1\\
        H'_{12}(1)&=&qS_{21}H'_{11}(1)+qS_{22}H'_{12}(1)+C_2
\end{eqnarray}
The mean cluster size diverges when
\begin{equation*}
    (1-qS_{11})(1-qS_{22})=q^2S_{12}S_{21}
\end{equation*}
The site-bond curve is given by - 
\begin{equation}
    q(p) = \frac{-S_{11}-S_{22}+\sqrt{(S_{11}+S_{22})^2+4(S_{12}S_{21}-S_{11}S_{22})}}{2(S_{12}S_{21}-S_{11}S_{22})}
\end{equation}
\begin{figure}
    \centering
    \includegraphics[scale=0.65]{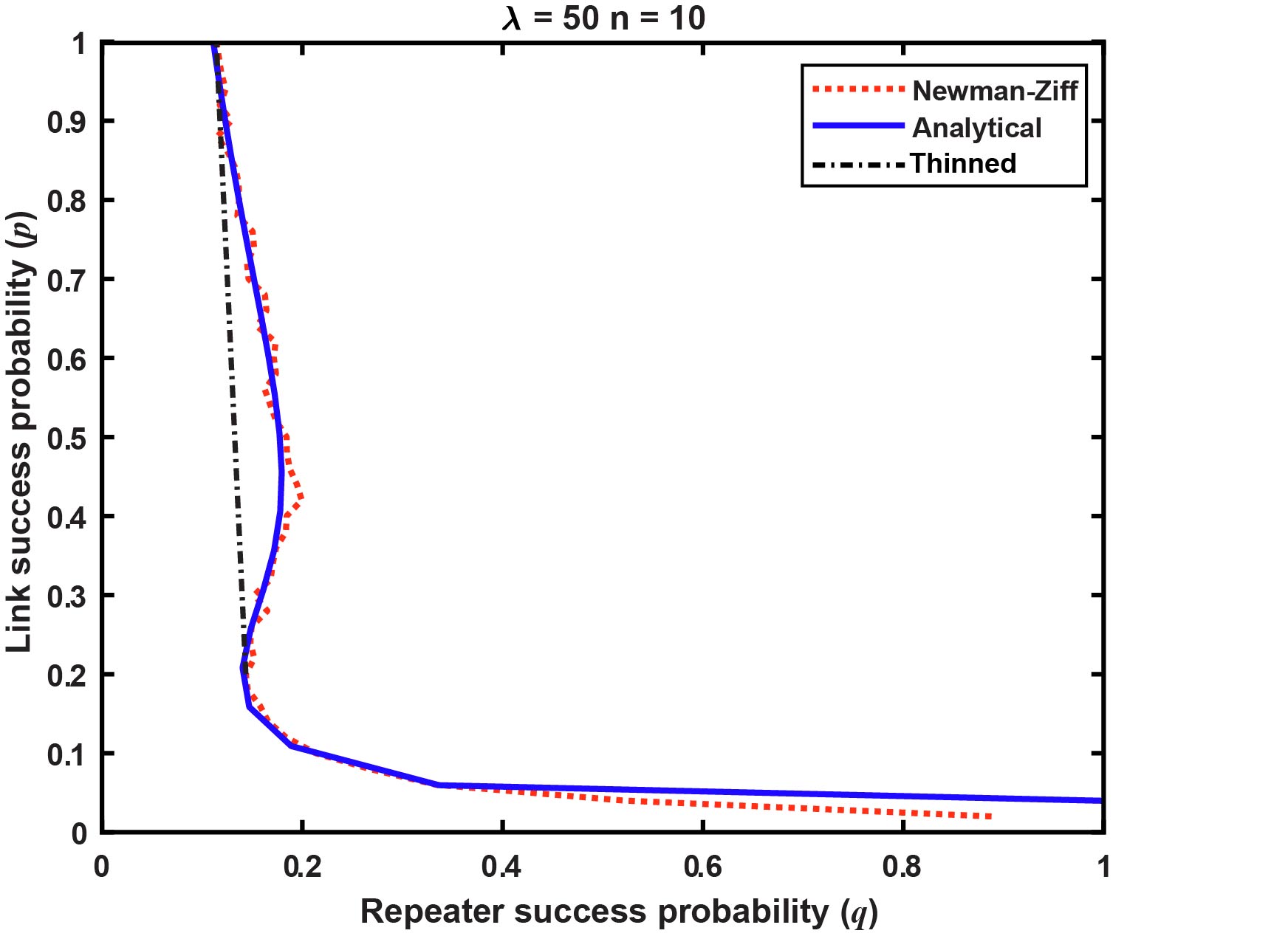}
    \caption{The site-bond region for Poisson-degree distributed random graph with mean node degree ($\lambda=50$) for 10-GHZ brickwork protocol}
    \label{fig:configuration_brickwork}
\end{figure}
\subsection{Rate calculation for 2-GHZ protocol}\label{app:2GHZ}
Consider 2-GHZ protocol on the square-grid network, i.e., the repeaters perform only Bell state measurements (BSMs) on the successful links. Let $d_{AB}$ be the Manhattan distance between Alice and Bob. For the link generation probability $p$, let $F(p)$ denote fraction of grid lying in giant connected component for a square lattice. Let the BSM success probability be $q$.  Then the shared entanglement generation rate $R$ is proportional to the probability that there exists a path between Alice and Bob in the graph generated after performing BSMs. For this protocol, the maximum possible achievable rate is 4 ebits/cycle. Hence, we can write,
\begin{equation}
    \label{eq:BSM_rate_decay}
    R \le 4F^2(p)q^{d_{AB}-1}
\end{equation}
This is a very loose upper bound on the rate. But it decays exponentially with the separation between Alice and Bob. Hence, it is impossible to achieve distance-independent rate by using only Bell state measurements.

\bibliography{bibFile.bib}

\begin{acknowledgments}
We thank Stefano Pirandola for useful discussions. AP and SG acknowledge the National Science Foundation (NSF) EFRI-ACQUIRE program, grant number ECCS-1640959. DT's work was supported in part by the NSF under grant CNS-1617437.
\end{acknowledgments}

\end{document}